\documentclass[twocolumn]{aastex62}
\usepackage{natbib}
\graphicspath{{./}{../pictures/}}

\received{}
\revised{}
\accepted{}
\submitjournal{ApJ}

\shorttitle{Tidal disruption events}
\shortauthors{Ryu et al.}

\usepackage{graphicx}	
\usepackage{amsmath}	
\usepackage{amssymb}	
\usepackage{cases}
\usepackage{array,multirow}
\usepackage{upgreek}
\usepackage{textcomp}

\usepackage{hyperref}
\usepackage[referable]{threeparttablex}
\usepackage{booktabs, dcolumn}
\makeatletter
\newcommand*{\rom}[1]{\expandafter\@slowromancap\romannumeral #1@}
\makeatother

\newcommand{\beq}{\begin{equation}}
\newcommand{\eeq}{\end{equation}}
\newcommand{\simlt}{\mathrel{\hbox{\rlap{\hbox{\lower4pt\hbox{$\sim$}}}\hbox{$<$}}}}
\newcommand{\simgt}{\mathrel{\hbox{\rlap{\hbox{\lower4pt\hbox{$\sim$}}}\hbox{$>$}}}}

\newcommand{\erg}{\;\mathrm{erg}}
\newcommand{\s}{\;\mathrm{s}}

\newcommand{\Msol}{\;\mathrm{M}_{\odot}}
\newcommand{\Rsol}{\;\mathrm{R}_{\odot}}

\newcommand{\km}{\;\mathrm{km}}

\newcommand{\yr}{\;\mathrm{yr}}

\newcommand{\rtidal}{r_{\rm t}}
\newcommand{\rg}{r_{\rm g}}
\newcommand{\physrad}{\mathcal{R}_{\rm t}}
\newcommand{\Mbh}{\left(\frac{M_{\rm BH}}{10^{6}}\right)}
\newcommand{\Mstar}{\;M_{\star}}
\newcommand{\Lphysr}{\mathcal{L}_{\rm t}}

\def\apjl{ApJL}
\def\apj{ApJ}
\def\mnras{M.N.R.A.S.}
\def\aap{A\&A}
\def\nat{Nat.}

\def\araa{Ann. Rev. A\&A}
\def\pasp{PASP}
\def\apjs{ApJ Supp.}

\def\physrep{Phys. Rep.}

\def\prd{prd}

\usepackage[normalem]{ulem}
\usepackage{booktabs,tikz}

\newcommand{\harm}{{\sc Harm3d}}   
\newcommand{\mesa}{{\small MESA}}

\defcitealias{Ryu1+2019}{Paper 1}
\defcitealias{Ryu2+2019}{Paper 2}
\defcitealias{Ryu3+2019}{Paper 3}
\defcitealias{Ryu4+2019}{Paper 4}
\begin{document}

\title{Tidal Disruptions of Main Sequence Stars - I. Observable Quantities and their Dependence on  Stellar and Black Hole Mass}

\correspondingauthor{Taeho Ryu}
\email{tryu2@jhu.edu}

\author[0000-0002-0786-7307]{Taeho Ryu}
\affil{Physics and Astronomy Department, Johns Hopkins University, Baltimore, MD 21218, USA}

\author{Julian Krolik}
\affiliation{Physics and Astronomy Department, Johns Hopkins University, Baltimore, MD 21218, USA}
\author{Tsvi Piran}
\affiliation{Racah Institute of Physics, Hebrew University, Jerusalem 91904, Israel}

\author{Scott C. Noble}
\affiliation{Gravitational Astrophysics Laboratory, Goddard Space Flight Center, Greenbelt, MD 20771, USA}

\begin{abstract}

This paper  introduces a series of papers presenting a quantitative theory for the tidal disruption of main sequence stars by supermassive black holes.  Using fully general relativistic hydrodynamics simulations and {\small{MESA}}-model initial conditions, we explore the pericenter-dependence of tidal disruption properties for eight stellar masses ($0.15\leq M_{\star}/\rm{M}_\odot\leq10$) and six black hole masses ($10^5\leq M_{\rm{BH}}/\rm{M}_\odot\leq5\times10^7$).
We present here the results most relevant to observations. The effects of internal stellar structure and relativity decouple for both the disruption cross section and the characteristic energy width of the debris. Moreover, the full disruption cross section is almost independent of $M_{\star}$ for $M_{\star}/\rm{M}_{\odot}\lesssim3$. Independent of $M_{\star}$, relativistic effects increase the critical pericenter distance for full disruption events by up to a factor $\sim 3$ relative to the Newtonian prediction. The probability of a direct capture is also independent of $M_{\star}$; at $M_{\rm{BH}}/\rm{M}_\odot\simeq5\times 10^6$ this probability is equal to the probability of a complete disruption. The breadth of the debris energy distribution $\Delta E$ can differ from the standard estimate by factors of $0.35-2$, depending on $M_{\star}$ and $M_{\rm{BH}}$, implying a corresponding change ($\propto(\Delta E)^{-3/2}$) in the characteristic mass-return timescale. 
 We provide analytic forms, suitable for use in both event rate estimates and parameter inference, to describe all these trends.  For partial disruptions, we find a nearly-universal relation between the star's angular momentum and the fraction of $M_{\star}$ remaining. Within the ``empty loss-cone" regime, partial disruptions must precede full disruptions. These partial disruptions can drastically affect the rate and appearance of subsequent total disruptions.

\end{abstract}


\section{Introduction} \label{sec:intro}

Supermassive black holes (SMBHs) reside in the nuclei of virtually every massive galaxy \citep{KormendyHo2013}. The orbits of stars around the central BH are stochastically perturbed by weak gravitational encounters with other stars.  Occasionally these perturbations place stars on orbits taking them so close to the BH that they are tidally disrupted, losing part or all of their mass in a tidal disruption event (TDE). Roughly half the mass torn off the star swings far out from the stellar orbit's pericenter and then returns. The energy it releases as it falls deeper into the black hole potential generates a luminous flare.

Many examples of tidal disruption events have now been seen.  Since the detection of the first TDE candidates \citep{KomossaBade1999} in the ROSAT all-sky survey \citep{Truemper1982}, greatly improved searches have been conducted, including X-ray surveys such as the XMM-$Newton$ slew survey \citep{Saxton+2008} and UV/optical surveys, e.g., the GALEX Deep Imaging Survey \citep{Gezari+2006}, Pan-STARRS \citep{Chambers+2016}, PTF \citep{Law+2009} and ASAS-SN \citep{Holoien+2016}.
From these, dozens of transients have been identified as TDE candidates \citep[e.g.,][]{Komossa2015,UV2018}. In the near future, this number is likely to grow rapidly with detections by ongoing surveys like the Zwicky Transient Facility \citep[ZTF,][]{Graham+2019} and upcoming surveys, e.g., the eROSITA All-Sky Survey \citep{Merloni+2012} and the Large Synoptic Survey Telescope \citep[LSST,][]{LSST2009}.

 An order-of-magnitude estimate for the ``tidal radius" see Equation~\ref{eq:tidalradius} below) is commonly used as an indicator of when a star is torn apart.  Another order-of-magnitude argument is used to estimate the energy spread of the debris in order to set the timescale of the event.  Newtonian dynamics underly both of these estimates, even though typically these events take place no more than a few tens of gravitational radii from a black hole.  Although tidal effects are strongly dependent upon distance from the star's center-of-mass, often no consideration is given to the stars' internal density profiles, or the polytrope approximation is taken to be general. The energy spread estimate is based upon conditions at a single point in the star's orbit (sometimes the ``tidal radius", sometimes the pericenter) although the star can travel a significant distance while it loses mass.  Finally, relatively little attention is paid to partial disruptions, although the rate at which they occur should be comparable to, or even larger than  to the rate of total disruptions. It is our goal to remove all these limitations.

To clarify the context in which we are working, it is useful to briefly expand upon the present state-of-the-art.  The term ``tidal radius" usually refers to an estimate of the pericenter distance at which tidal effects can be important to stars \citep{Hills1988}:
\begin{align}
\label{eq:tidalradius}
\rtidal&=\left(\frac{M_{\rm BH}}{M_{\star}}\right)^{1/3}R_{\star}\\\nonumber
&\simeq 47\left(\frac{M_{\rm BH}}{10^{6}\Msol}\right)^{-2/3}\left(\frac{M_{\star}}{1\Msol}\right)^{-1/3}\left(\frac{R_{\star}}{1\Rsol}\right)r_{\rm g},
\end{align}
where $M_{\star}$ and $R_{\star}$ are the stellar mass and radius, respectively. $M_{\rm BH}$ is the mass of the BH and $r_{\rm g}$ is the gravitational radius of the BH, with $r_{\rm g} \equiv GM_{\rm BH}/c^{2}$.  Following \citet{Rees1988}, it is generally assumed that the energy distribution of the debris mass $dM/dE$ is  non-zero for $-\Delta \epsilon \leq E \leq +\Delta \epsilon$. The characteristic energy spread $\Delta\epsilon$ is set to an order-of-magnitude estimate for the range in fluid binding energies \citep{Stone+2013}
\begin{align}\label{eq:deltae}
\Delta \epsilon &= \frac{GM_{\rm BH}R_{\star}}{r_{\rm t}^{2}}= \left(\frac{R_{\star}}{r_{\rm g}}\right)\left(\frac{r_{\rm t}}{r_{\rm g}}\right)^{-2}c^2, \nonumber\\
&\sim 2 \times 10^{-4} \left(\frac{R_{\star}}{\rm{R}_\odot}\right)^{-1} \left(\frac{M_{\rm BH}}{10^{6}\Msol}\right)^{1/3} \left(\frac{M_{\star}}{1\Msol}\right)^{2/3} c^2.
\end{align}
Sometimes $r_{\rm t}$ is replaced with the actual pericenter of the orbit $r_{\rm p}$ \citep[e.g.,][]{Lodato+2009}.  

\citet{Phinney1989} was the first to recognize that $\rtidal$ is not exactly the maximum orbital pericenter for a complete tidal disruption, a distance we would like to name the ``physical tidal radius" (we assign it the symbol $\physrad$). To remedy the neglect of internal stellar structure, he suggested that $\physrad$ could be estimated by applying to $\rtidal$ a correction factor based on the star's apsidal motion constant and its dimensionless binding energy.  For this reason, $r_{\rm t}$ is sometimes reinterpreted to be ${\cal R}_{\rm t}$, but without evaluating how it might differ from $\rtidal$ \citep{Stone+2013}. Several groups have tried to include stellar structure in the calculation of $\physrad$, but employing purely Newtonian dynamics on polytropic stars  \citep[e.g.,][]{LuminetCarter1986,Khokhlov+1993,Guillochon+2013,Mainetti+2017}. Recently, there have been efforts beginning from genuine main sequence stellar structures, but still restricted to Newtonian dynamics, and examining a limited range of stellar masses (only $1\Msol$ in \citealt{Goicovic+2019}, $1\Msol$ and $3\Msol$ at several ages in \citealt{LawSmith+2019}, $0.3\Msol$, $1\Msol$ and $3\Msol$ at three different ages in \citealt{Golightly2+2019}). Others have explored the dependence on black hole mass induced by relativistic effects, but without any reference to internal stellar structure or the hydrodynamics of disruption \citep[e.g.,][]{IvanovChernyakova2006,Kesden2012,ServinKesden2017}.
Earlier works  employed a post-Newtonain approximation \citep{Ayal_2000} or explored the use of relativistic hydrodynamics simulations for strong encounters of polytropic stars, but without stellar self-gravity \citep[e.g.,][]{Laguna+1993}.  In some cases,  relativistic effects were approximated by a ``generalized Newtonian potential" \citep{Gafton2015,Gafton2019} or in terms of genuine relativistic dynamics \citep{Frolov+1994}, but assuming a polytropic structure for the star and computing stellar self-gravity in an entirely Newtonian fashion (in the last case, fixing it to its initial stellar surface value).  Many of these explorations of $\physrad$ also computed the energy distribution $dM/dE$ and explored the relation between the remnant mass and orbital pericenter in partial disruptions; in one case \citep{Manukian+2013}, they also examined the remnants' orbital properties. However, all this work was subject to the limitations already enumerated.

\begin{figure*}
	\centering
\includegraphics[width=14cm]{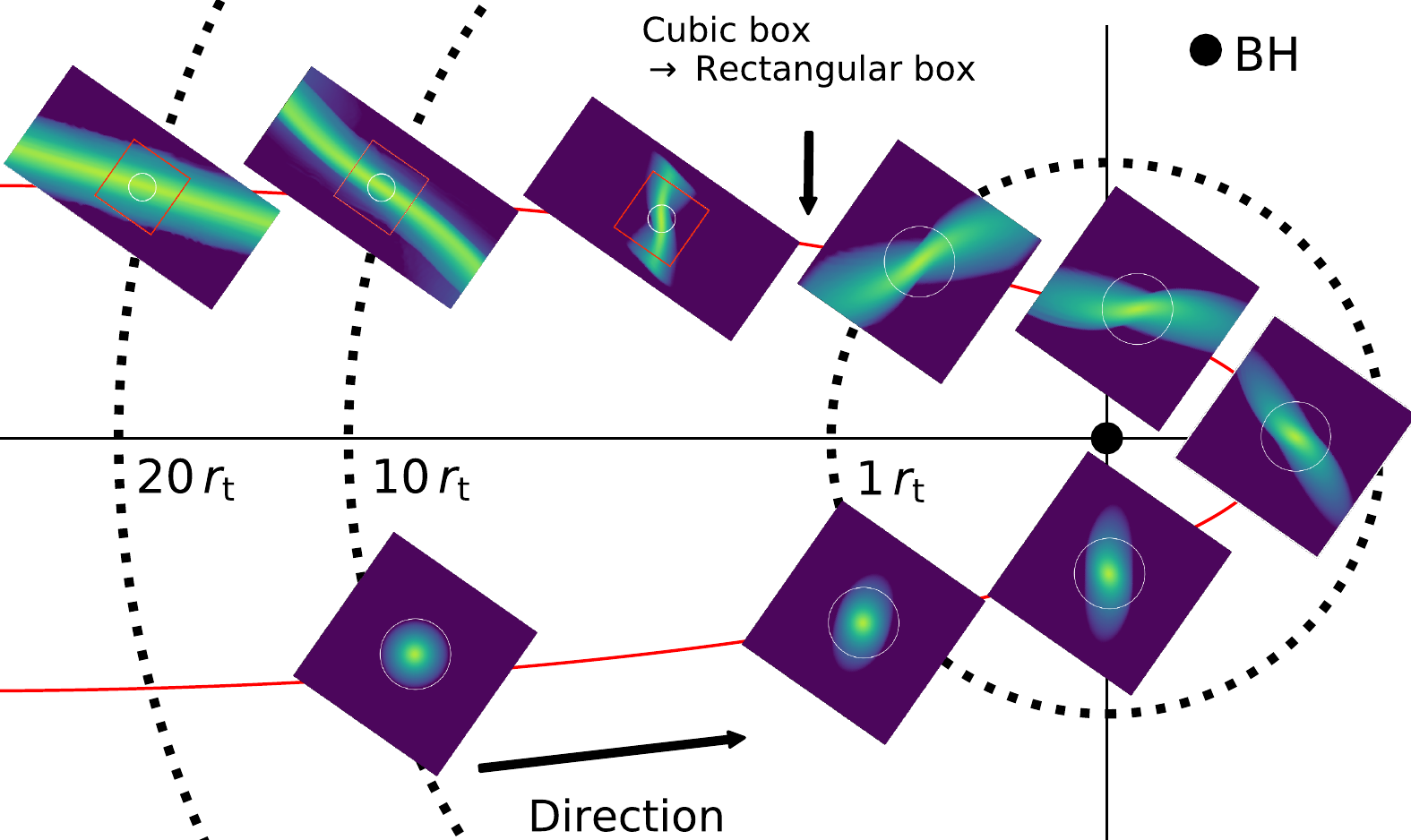}
\caption{Successive moments in a full TDE. The red line indicates the star's orbit around the black hole (black circle). Each inset figure presents a snapshot of the density distribution in the orbital plane within our simulation box.
The white circle in each snapshot shows the initial stellar radius. Partway through the event, we replace the cubic box with a rectangular box (see \citetalias{Ryu2+2019}); we  draw a red square in the rectangular boxes to show the position and size of the original cubic box. Note that the rectangular boxes are not drawn to the same scale as the cubic boxes, and the dotted curves marking $\rtidal$, $10~\rtidal$ and $20~\rtidal$ are likewise not drawn to scale.}
\label{fig:overview}
\end{figure*}

This is the first in a series of four papers in which we present the results of a large number of simulations designed to include all relevant physical processes.  Tidal stresses are treated in exact general relativity, as are the fluid dynamics of the disrupted star.  The stars' initial states are taken from the stellar evolution code \mesa, with ages halfway through their main sequence lifetimes so as to correspond to the time-averaged state of main-sequence disruptions.  Stellar self-gravity is computed with the Newtonian Poisson equation, but in a tetrad frame comoving with the star whose metric (within the simulation volume) departs from Minkowski only by very small amounts (see Appendix in \citealt{Ryu2+2019}). For a fiducial black hole mass of $10^6 \Msol$, we treat stars of eight different masses $M_{\star}$, from $0.15\Msol$ to $10\Msol$.  For three of these stellar masses ($0.3\Msol$, $1\Msol$, and $3\Msol$), we consider black holes of six different masses, from $10^5 \Msol$ to $5\times 10^7\Msol$.  All our black holes, however, have zero spin.  In each simulation, the star's trajectory has eccentricity $e$ such that $1-e \simeq 10^{-8}$. Both to closely determine $\physrad$ and to map out how the properties of partial disruptions depend on $r_{\rm p}/\physrad$, we simulated encounters for each $(M_{\star},M_{\rm BH})$ pair for a number of pericenters spaced by $\simeq 0.05-0.2~\rtidal$.

\begin{figure*}
	\centering
	\includegraphics[width=8.9cm]{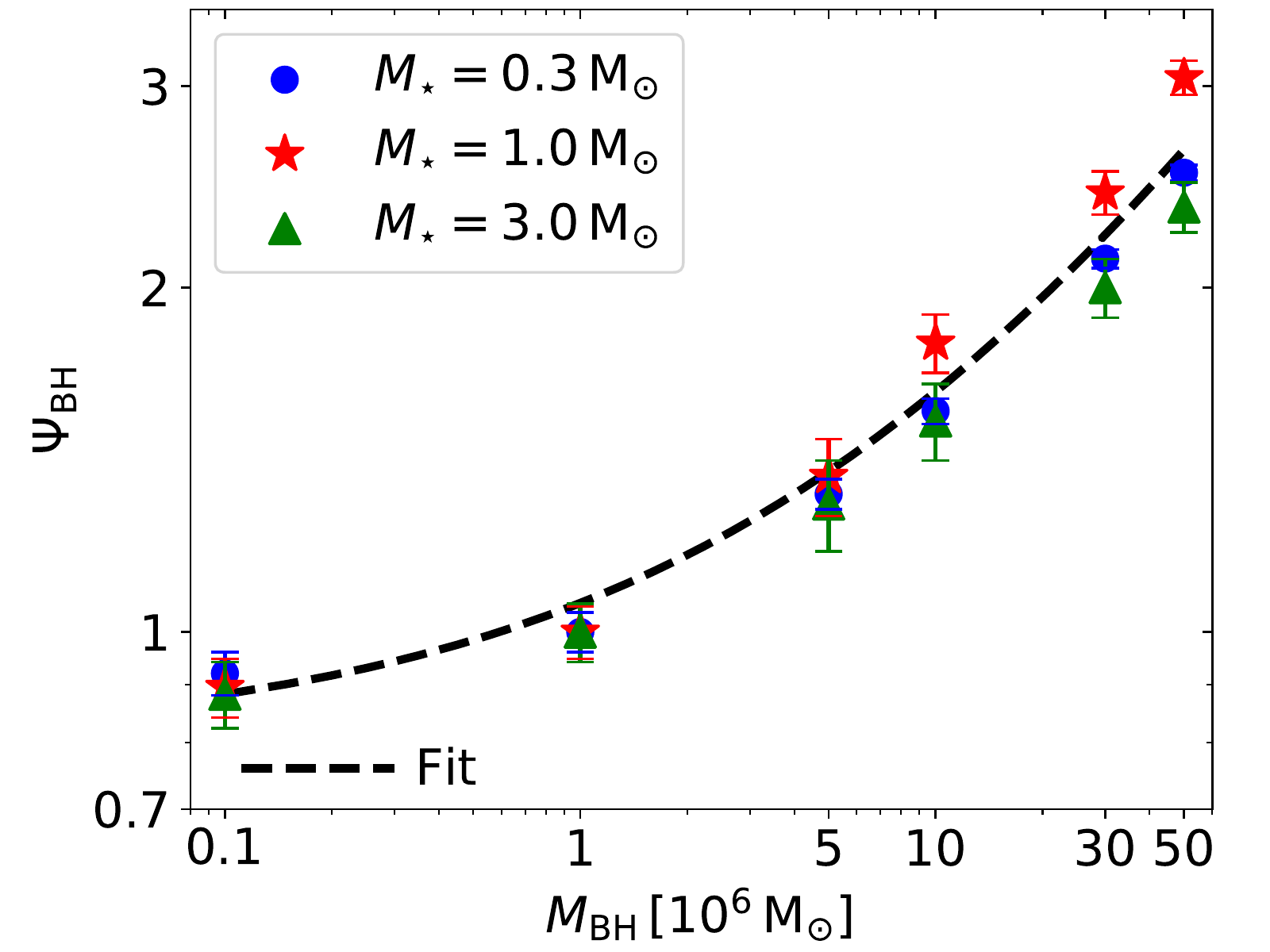}		
	\includegraphics[width=8.9cm]{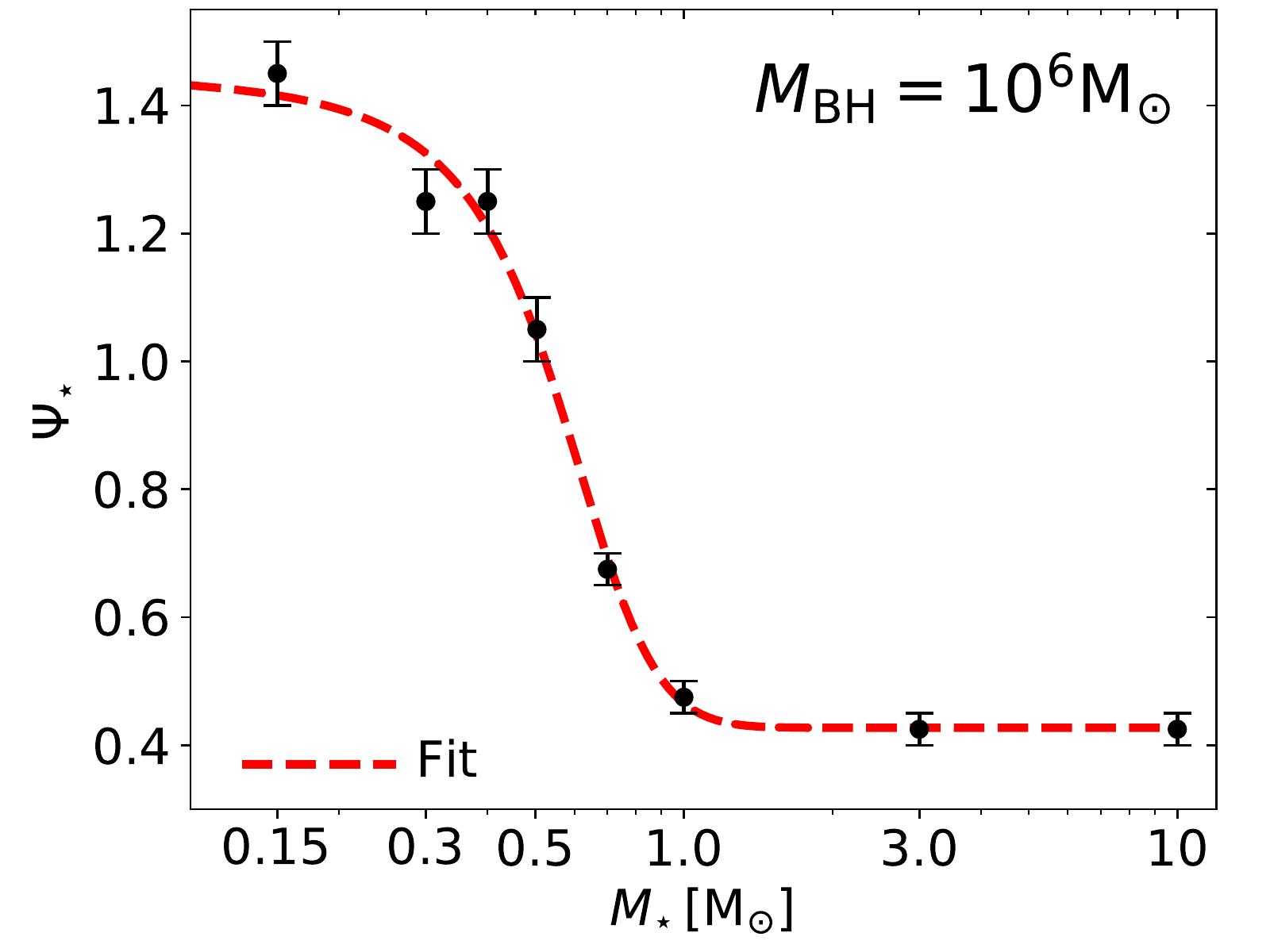}	
	\vspace{-0.1in}
	\caption{(\textit{Left} panel) $ \Psi_{\rm BH} \equiv \Psi (M_{\rm BH}; M_{\star})/ \Psi(10^6;M_{\star})$ 
	as a function of $M_{\rm BH}$ for $M_{\star}=0.3$ (blue circles), $1$ (red stars) and $3$ (green triangles). $\Psi$ is defined as $\mathcal{R}_{\rm t}/r_{\rm t}$.
 The dashed line depicts our fit to $\Psi_{\rm BH}$ (Equation \ref{eq:fit_psibh}
		(\textit{Right} panel) $\Psi(M_\star,10^6) \equiv \Psi_\star(M_\star) \equiv \physrad/r_{\rm t}$ for $M_{\rm BH}=10^{6}$. Numerical values are shown by filled circles, the analytic fit given in Equation~\ref{eq:fit_psistar} by a red dashed curve. In both panels, the error bars indicate the uncertainties in $\physrad$ originating from the finite sampling of $r_{\rm p}/r_{\rm t}$.
		}
	\label{fig:r_tidal_bh}
\end{figure*}

A schematic overview of the entire process can be viewed in Figure~\ref{fig:overview}. The star begins to be stretched when its distance to the black hole is $\rtidal$, and is already quite distorted by the time it reaches its pericenter (here $\sim \rtidal/2$).   However, substantial mass-loss continues until the star has traveled far from the black hole (see \citealt{Ryu2+2019} for full details.)

In this first paper, we give an overview of this series' principal findings: the physical tidal radius as a function of stellar mass and black hole mass (Section \ref{sec:physicalR}), the full disruption cross section (Section~\ref{subsub:tde_mstar}), the energy scale of stellar debris from full disruptions and its scaling with $M_{\star}$ and $M_{\rm BH}$ (Section \ref{sec:energyscale}) and the relationship between remnant mass and orbital angular momentum in partial disruptions (Section \ref{sec:partial}). We then discuss the implications these results have for predicted event rates for full and partial disruptions (Section \ref{sub:r_tderate}), and for the orbital properties of stellar debris from full disruptions (Section \ref{subsub:debris}).  
We conclude  with a summary of our findings (Section~\ref{sec:summary}).

The following three papers in this series provide details supporting and expanding upon the findings discussed in this paper. In \citet{Ryu2+2019} (\citetalias{Ryu2+2019}) and \citet{Ryu3+2019} (\citetalias{Ryu3+2019}), we focus on the stellar mass dependence of disruption outcomes: In \citetalias{Ryu2+2019}, we: describe  our methodology, including hydrodynamic algorithms, relativistic self-gravity calculation, and grid-resolution; explain our \mesa -based initial conditions and simulation setup; present our detailed results having to do full disruptions; and compare these results to related work by others. \citetalias{Ryu3+2019} reports our results relevant to partial disruptions. In \citet{Ryu4+2019} (\citetalias{Ryu4+2019}), we explore how relativistic effects lead to dependence of TDE properties on black hole mass.   There we demonstrate how our study extends prior efforts concerning relativistic effects and evaluate the quality of various approximations to general relativistic physics in the TDE literature.

Hereafter all masses are measured in solar mass and stellar radii in solar radius.

\section{Physical tidal distance $\physrad$ }
\label{sec:physicalR_crossection}
\subsection{Physical tidal distance $\physrad$}
\label{sec:physicalR}
\begin{figure*}
	\centering
		\includegraphics[width=8.9cm]{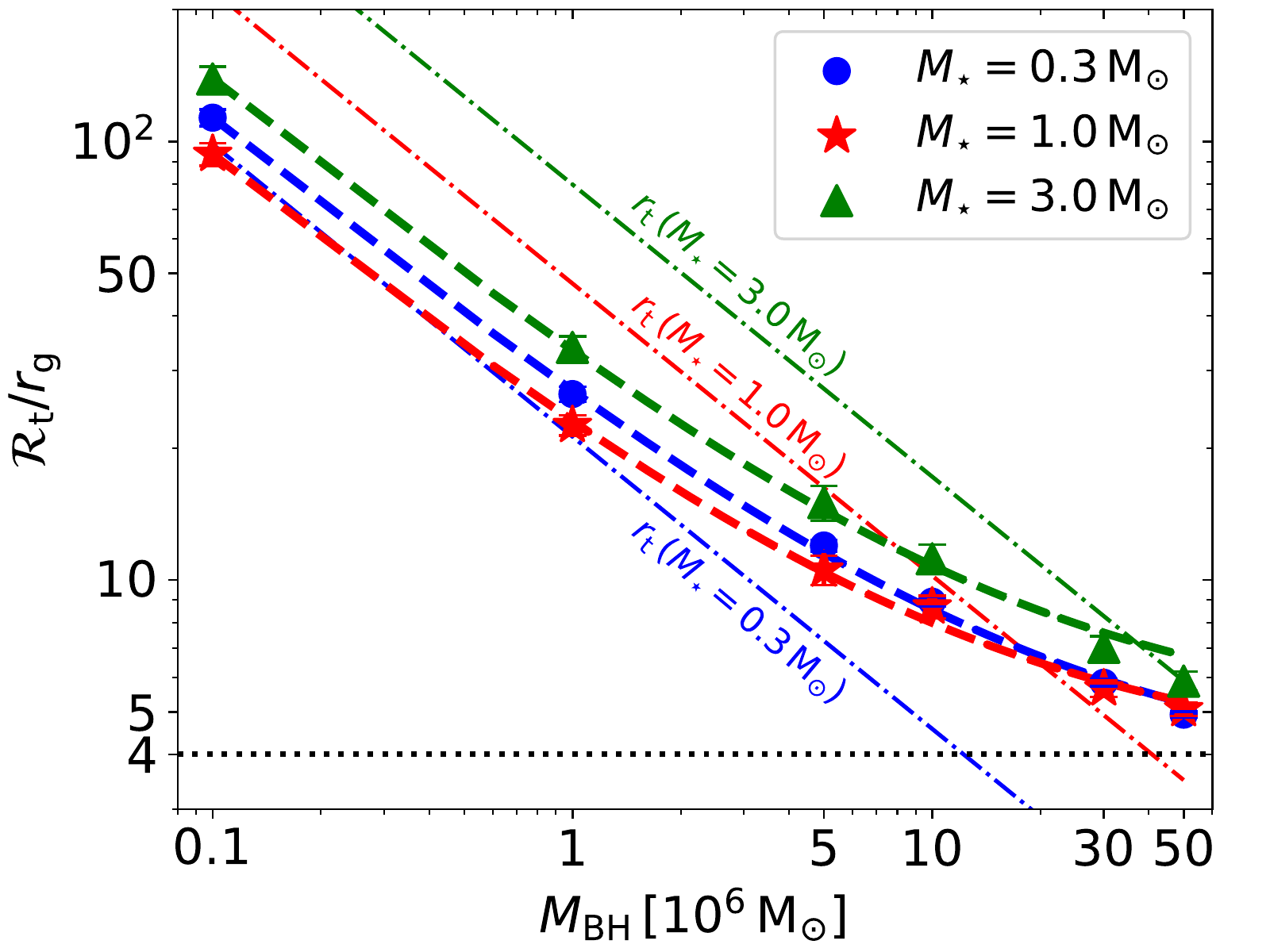}
	\includegraphics[width=8.9cm]{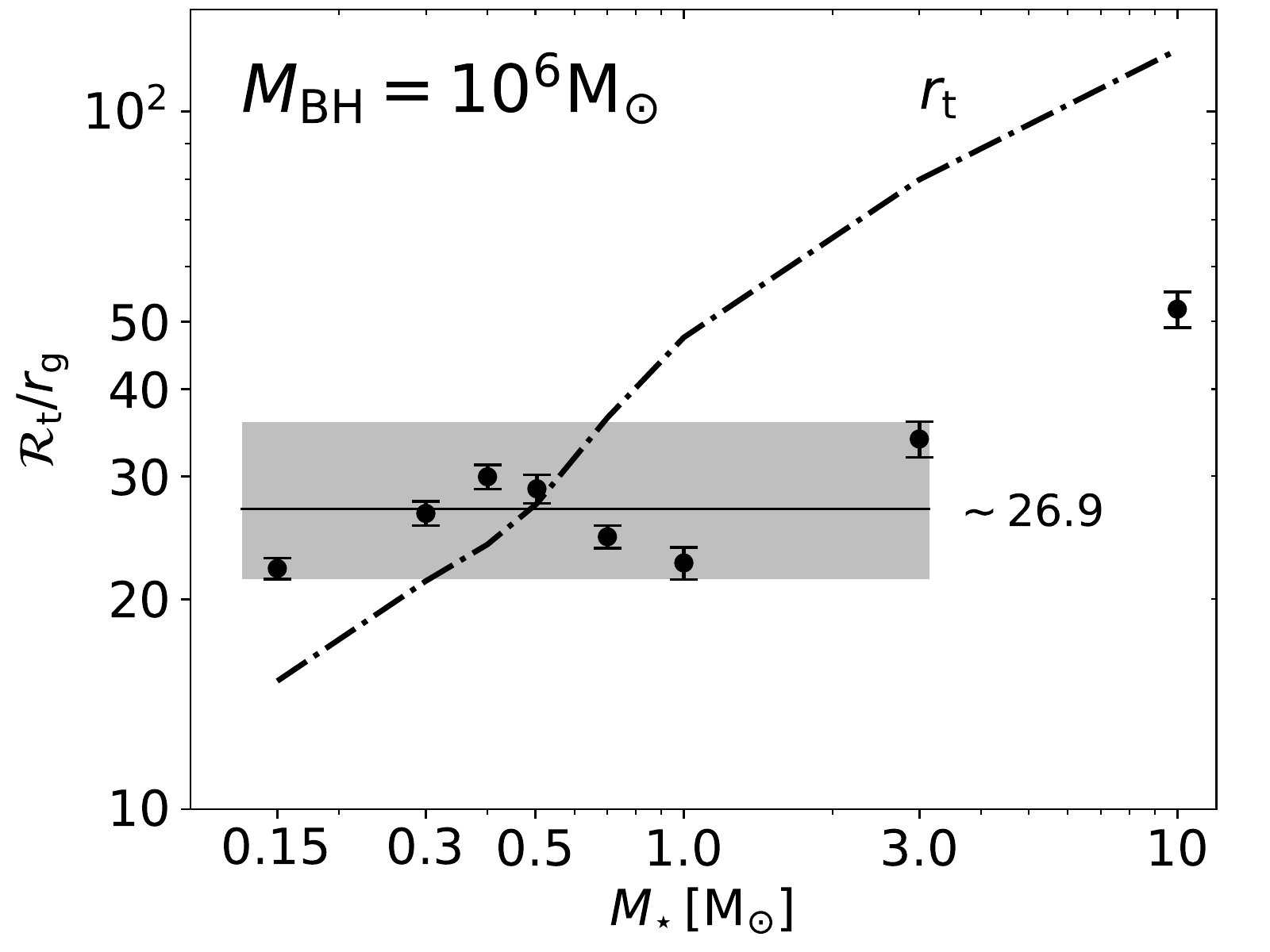}
		\vspace{-0.1in}
	\caption{$\mathcal{R}_{\rm t}/\rg$.	(\textit{Left} panel) $\physrad/\rg$ as a function of $M_{\rm BH}$ for $M_{\star}=0.3$ (blue circles), $1$ (red stars) and $3$ (green triangles). The dashed curves show the fitting formulae given in Equation~\ref{eq:fit_R_t_BH}. The diagonal dash-dot lines represent the order-of-magnitude estimates of tidal radius $\rtidal$ (Equation~\ref{eq:tidalradius}) for the stellar masses. The dotted horizontal line indicates $\physrad/\rg=4$, which is the minimum pericenter distance for parabolic (non-plunging) orbits in Schwarzschild spacetime.  (\textit{Right} panel) $\physrad/r_{\rm g}$ as a function of $M_{\star}$ for $M_{\rm BH}=10^{6}$. Simulation results are shown by filled circles, while the traditional estimate $r_{\rm t}$ is represented by a dash-dot curve. The mean value and extreme range of $\physrad/r_{\rm g}$ for $0.15 \leq M_{\star} \leq 3$ are marked by a horizontal line and a shaded region, respectively. }
	\label{fig:r_tidal_star}
\end{figure*}

The physical tidal radius $\physrad$, the maximum pericenter for a total disruption, plays a key role in determining the fate of a TDE.  We define $\Psi$ as the ratio of  $\physrad$ to the ``order of magnitude tidal radius" $\rtidal$ defined in Equation~\ref{eq:tidalradius}, i.e., $\Psi\equiv\physrad/\rtidal$.
The physical radius $\physrad$ depends on both stellar mass $M_{\star}$ and black hole mass $M_{\rm BH}$.   Combining results from \citetalias{Ryu2+2019} and \citetalias{Ryu4+2019}, we find, as illustrated in the \textit{left} panel of Figure~\ref{fig:r_tidal_bh}, that the $M_{\rm BH}$-dependences for different stellar masses are essentially identical. This fact allows us to factor out the dependence of $\Psi$ on $M_{\star}$ from its dependence on $M_{\rm BH}$,
\begin{align}
    \Psi(M_{\rm BH},M_{\star}) \equiv\frac{\physrad}{\rtidal} = \Psi_{\rm BH}(M_{\rm BH})\Psi_{\star} (M_{\star}),
\end{align}
where $\Psi_{\star}$ accounts for the stellar internal structure while $\Psi_{\rm BH}$ encapsulates the behavior due to relativistic effects. We choose to describe these two functions with $\Psi_{\rm BH}$ normalized to unity for $M_{\rm BH}=10^6$.

\begin{figure}
	\centering
	\includegraphics[width=9.0cm]{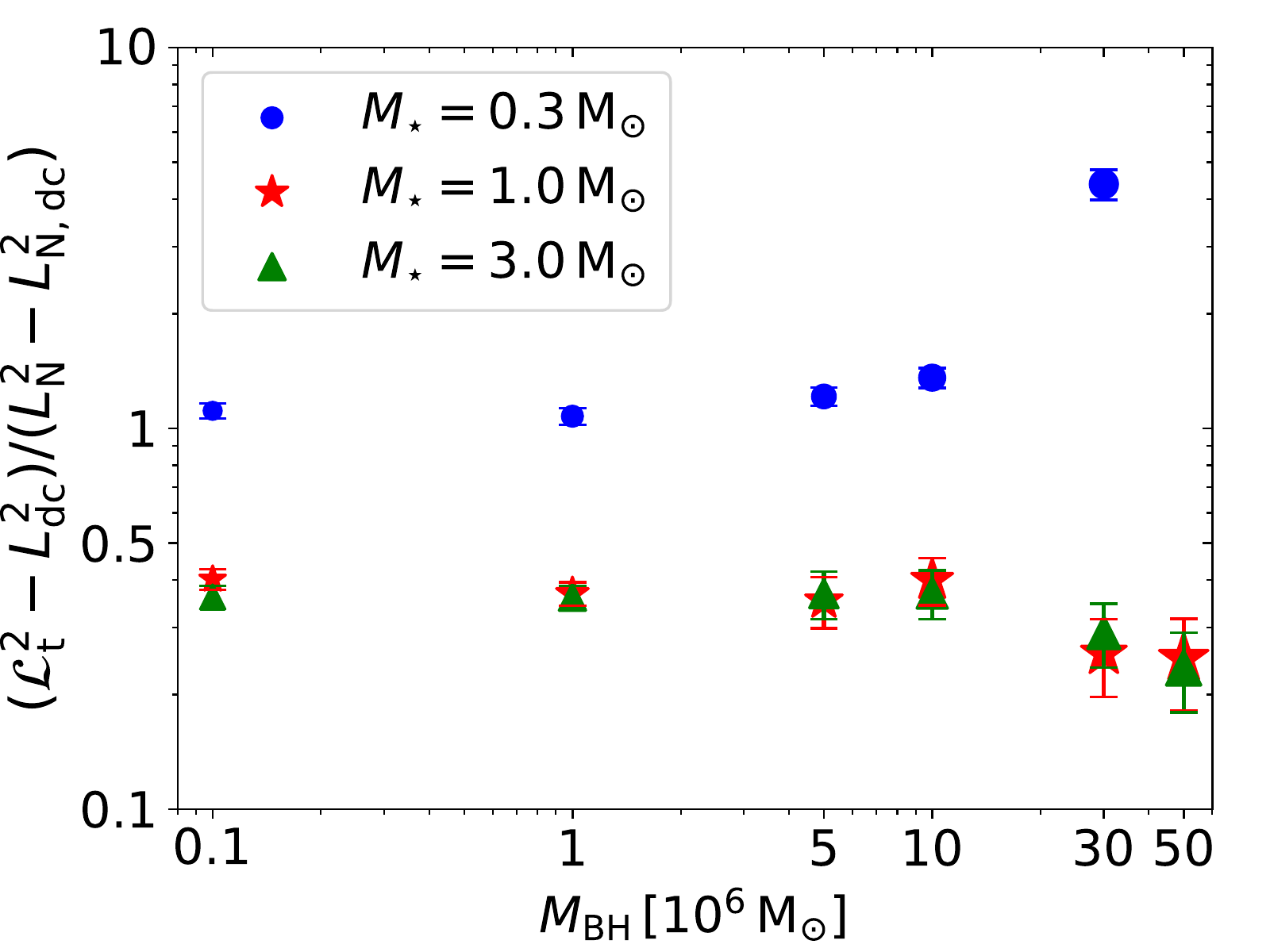}
	\caption{The ratio of the relativistic full tidal disruption cross section to its Newtonian analog, $(\Lphysr^2 - L_{\rm dc}^2)/(L_{\rm N}^2-L_{\rm N,dc}^2)$, as a function of $M_{\rm BH}$ for three stellar masses.The error bars indicate the errors propagated from the uncertainties of $\mathcal{R}_{\rm t}$.
Note that the ratio for $M_{\star}=3$ (blue circles) and $M_{\rm BH}=5\times10^{7}$ is not plotted because $L_{\rm N}<L_{\rm N,dc}$.}
	\label{fig:com_NT_GR}
\end{figure}

For all three values of $M_{\star}$ we studied (i.e., $M_{\star}=0.3$, 1, and 3) the dependence of $\Psi$ on $M_{\rm BH}$ is very well described by
\begin{equation}
\Psi_{\rm BH}(M_{\rm BH})=0.80 + 0.26~\left(\frac{M_{\rm BH}}{10^{6}}\right)^{0.5},\label{eq:fit_psibh}
\end{equation}
as depicted by the dashed curve in the \textit{left} panel of Figure~\ref{fig:r_tidal_bh}.  From the Newtonian limit (represented by $M_{\rm BH}=10^5$) to the highest black hole mass we examined ($M_{\rm BH} = 5 \times 10^7$), $\Psi_{\rm BH}$ increases by a factor of $3$, with its slope continually steepening as $M_{\rm BH}$ grows.

Just as the dashed curve in the \textit{left} panel of Figure~\ref{fig:r_tidal_bh} represents the black hole mass-dependence of $\physrad$ for all stellar masses, the dashed curve in the \textit{right} panel represents the shape of the stellar mass dependence of $\Psi$ ($\Psi_{\star}$) for all black hole masses; it is well-described by the expression
\begin{align}
\Psi_{\star}(M_{\star})& = \frac{1.47+ ~\exp[(M_{\star}-0.669 )/0.137]}{1 + 2.34~\exp[(M_{\star}-0.669)/0.137]}.
\label{eq:fit_psistar}
\end{align}
Although $\Psi_{\star}$ becomes nearly constant at both mass extremes, it has a sharp transition across the range of masses $0.4\lesssim M_{\star} \lesssim 1$.  For low-mass stars
\footnote{For explanatory convenience, we categorize stars into ``low-mass'' ($M_{\star}\leq0.5$) and ``high-mass'' ($M_{\star}\geq1)$ based on the properties of TDE outcomes.  These mass ranges may be different from those typically used in stellar evolution studies.}
($M_{\star} \leq 0.5$), which are predominantly convective, $\Psi_{\star} \simeq 1-1.45$. For higher mass stars ($M_{\star} \geq 1$), which are predominantly radiative, $\Psi_{\star} \simeq 0.45$.  The large coefficient of $M_{\star}$ in the exponentials of Equation~\ref{eq:fit_psistar} conveys how sharp the transition is from low-mass to high-mass stars.

When $\physrad$ is measured not in ratio to $\rtidal$, but in physical units (e.g., $\rg=GM_{\rm BH}/c^{2}$), its value depends on $M_{\rm BH}$, but is {\it nearly independent of $M_{\star}$} over most of the (logarithmic) range of possible stellar masses.  For $M_{\rm BH} = 10^6$, it is almost constant at $\simeq ~27~\rg$ from $M_{\star}\simeq0.15$ to $M_{\star} \simeq 3$ (the \textit{right} panel of Figure~\ref{fig:r_tidal_star}), with a maximum departure only 20\% either up or down. Such a small contrast in $\physrad$ across nearly the entire range of stellar mass is very different from the strong dependence on $M_{\star}$ predicted by the order-of-magnitude estimate $\rtidal$ (the dot-dashed curve in the \textit{right} panel of Figure~\ref{fig:r_tidal_star}), which rises as $\propto M_{\star}^{-1/3}R_{\star}\propto M_{\star}^{0.55}$ (\citetalias{Ryu2+2019}). The fact that $\physrad/r_{\rm g}$ is nearly independent of stellar mass likely also explains the near $M_{\star}$-independence of $\Psi_{\rm BH}$: $\physrad$ lies at roughly the same location in the black hole potential for stars of all masses.

The value of $\mathcal{R}_{\rm t}$ is primarily determined by the star's central density  $\rho_{\rm c}$ rather than by its mean density $\bar{\rho}_{\star}(=3M_{\star}/[4\uppi R_{\star}^{3}])$.  In  \citetalias{Ryu2+2019} we demonstrate that
\begin{align}
\mathcal{R}_{\rm t} \simeq 2.13\left(\frac{\bar{\rho}_{\star}}{\rho_{\rm c}}\right)^{1/3} \rtidal
\simeq 1.32\left(\frac{M_{\rm BH}}{\rho_{\rm c}}\right)^{1/3}.
\label{eq:Rt-zeta}
\end{align}

The increase in $\Psi_{\rm BH}$ toward larger $M_{\rm BH}$ results in a slower decrease in $\physrad/\rg$ with $M_{\rm BH}$ than would be predicted by Newtonian dynamics (the \textit{left} panel  of Figure~\ref{fig:r_tidal_star}). Whereas the Newtonian prediction is that $\rtidal/\rg \propto M_{\rm BH}^{-2/3}$, for $M_{\rm BH}\gtrsim5\times10^{6}$, $-d\ln (\physrad/\rg)/d \ln M_{\rm BH} \simeq 0.3$--0.4.  Nonetheless, the $M_{\rm BH}$-dependence of $\physrad/\rg$ can still be well-described with a term $\propto M_{\rm BH}^{-2/3}$, provided a weakly $M_{\star}$-dependent offset is added:
\begin{align}\label{eq:fit_R_t_BH}
\frac{\mathcal{R}_{\rm t}}{r_{\rm g}} =\begin{cases}
23.5 ~\left(\frac{M_{\rm BH}}{10^{6}} \right)^{-2/3} + 3.5 \hspace{0.1in} \text{for $M_{\star}=0.3$,}\\
19.1 ~\left(\frac{M_{\rm BH}} {10^{6}}\right)^{-2/3} + 3.9 \hspace{0.1in} \text{for $M_{\star}=1.0$,}\\
28.8 ~\left(\frac{M_{\rm BH}}{10^{6}} \right)^{-2/3} + 4.6\hspace{0.1in} \text{for $M_{\star}=3.0$.}
\end{cases}
\end{align}

\subsection{Event cross sections: comparison with estimates based on $r_{\rm t}$.} \label{subsub:tde_mstar}

\begin{figure}
	\centering
		\includegraphics[width=8.9cm]{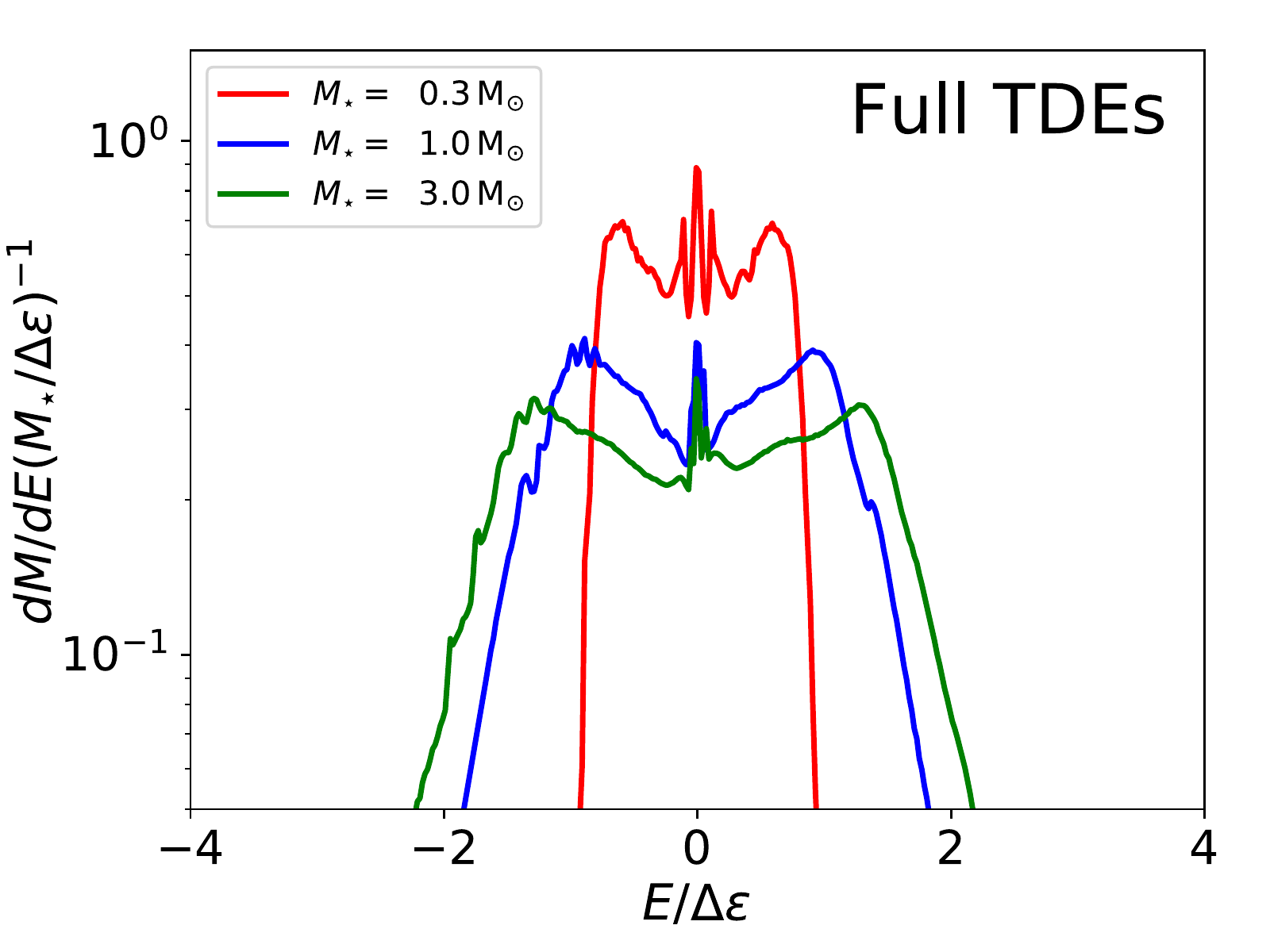}
	\caption{ The energy distribution $dM/dE$ of debris (in units of $M_{\star}/\Delta\epsilon$) for stars with $M_{\star}=0.3$ (red), 1 (blue) and 3 (green) fully disrupted by a $10^{6}\Msol$ black hole. The cases are the strongest encounters considered in this study for the given masses.}
	\label{fig:dmde}
\end{figure}

\begin{figure*}
	\centering
	\includegraphics[width=8.9cm]{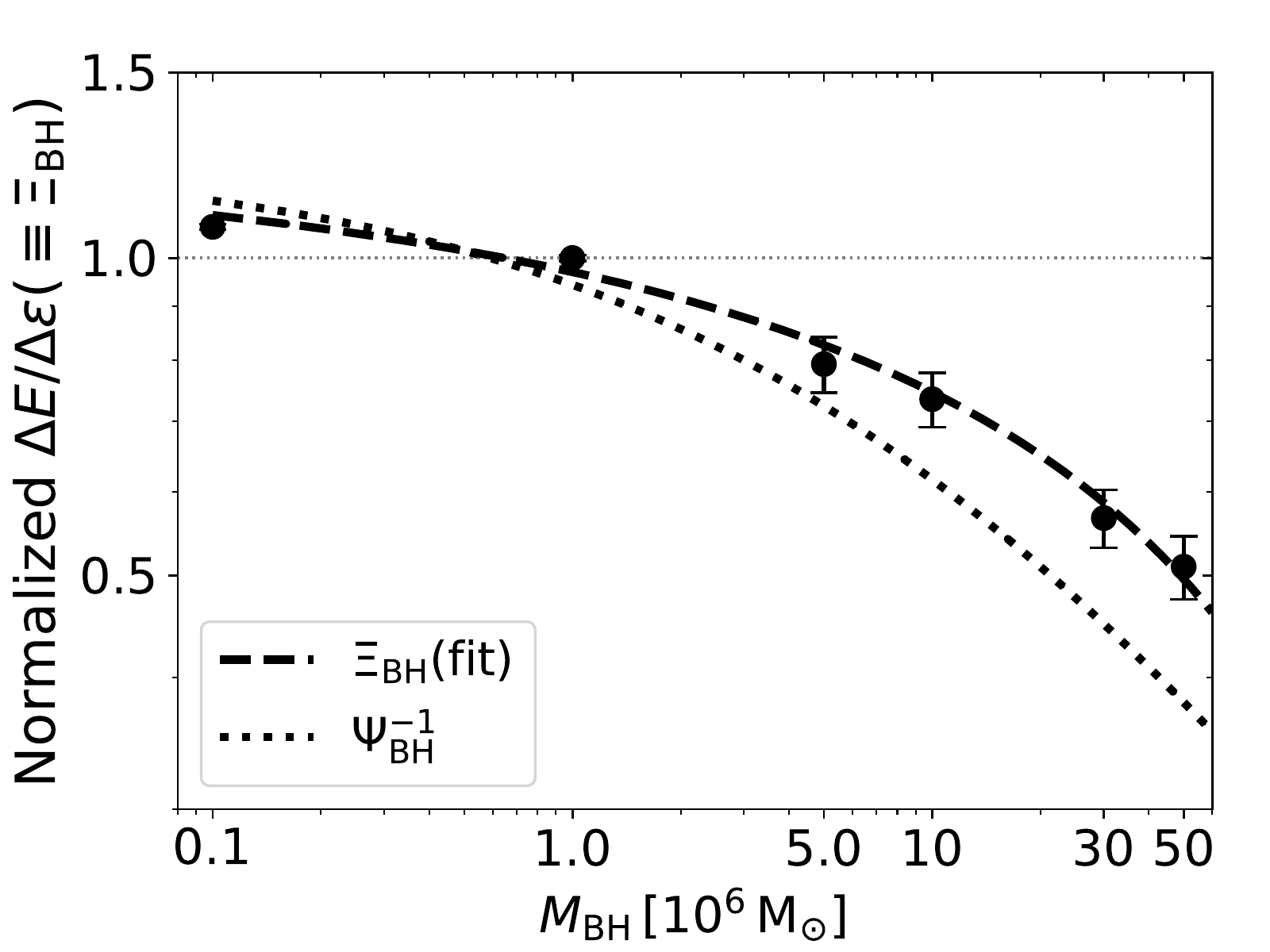}	
	\includegraphics[width=8.9cm]{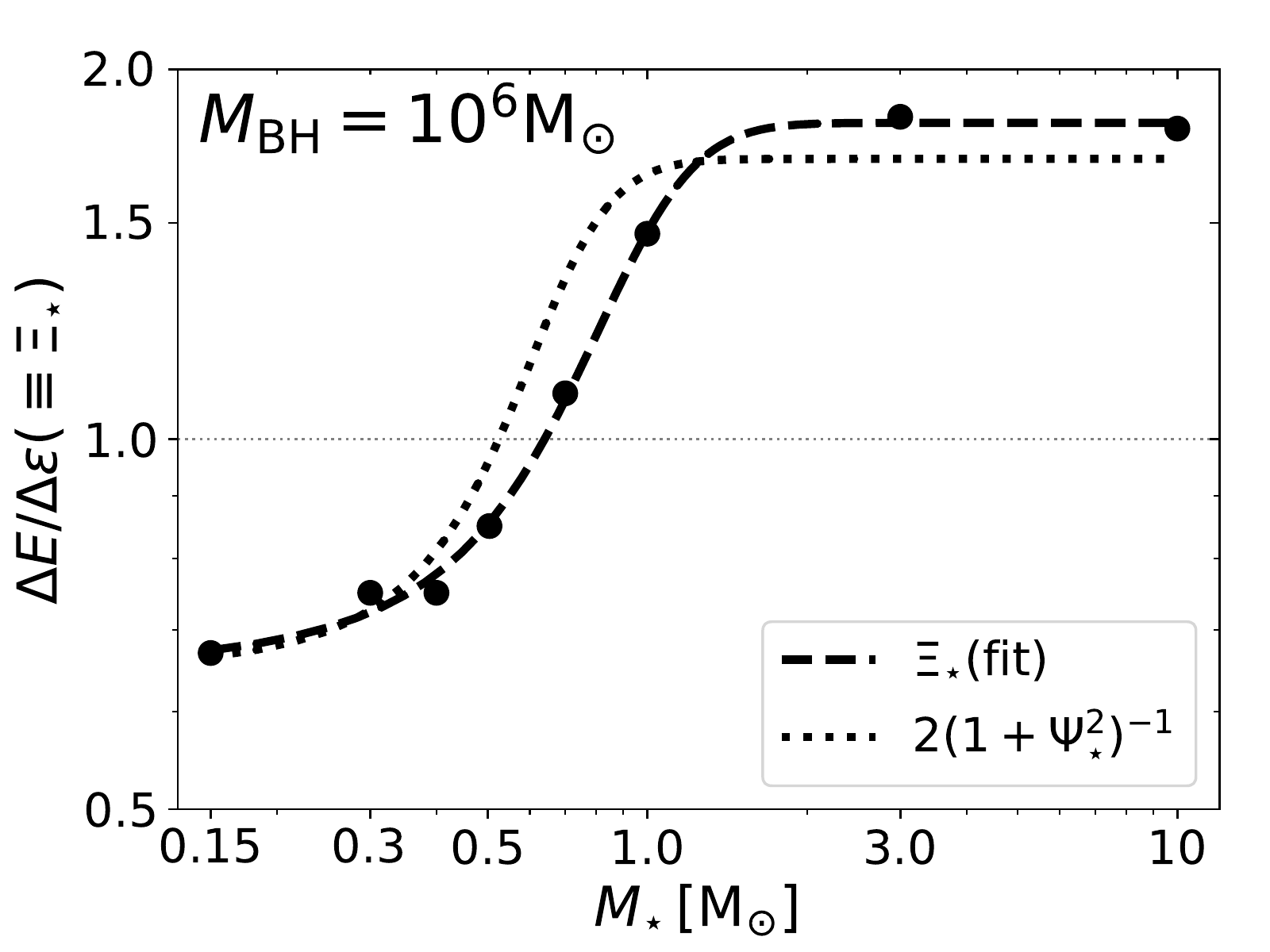}
	\caption{The characteristic debris energy spread $\Delta E$ in units of $\Delta\epsilon$. (\textit{Left} panel) $\Delta E/\Delta\epsilon$ normalized so that it is unity for $M_{\rm BH} = 10^6$.  The filled circles show the average over all the pericenters yielding full disruptions for three stellar masses ($M_{\star}=0.3$, $1$ and $3$); the error bars indicate the standard deviation with respect to stellar mass.
	(\textit{Right} panel) $ \Delta E/\Delta\epsilon$ for $M_{\rm BH}=10^{6}$ as a function of $M_{\star}$.
	In both panels, the dashed curves show the fitting formulae given in Equation~\ref{eq:Xi_BH} for $\Xi_{\rm BH}$ (\textit{left} panel) and Equation~\ref{eq:Xi_star} for $\Xi_{\star}$ (\textit{right} panel). The dotted curves show the approximation given in Equation~\ref{eq:energy_fit}.
}
	\label{fig:dmde_maxrange}
\end{figure*}

The cross section for tidal encounters with pericenter $\leq r_{\rm p}$ is proportional to $L^{2}(r_{\rm p})$. Here, $L$ is the specific angular momentum of the star's orbit. For parabolic orbits, $L$ is related to $r_{\rm p}$ by
\begin{equation}\label{eq:L}
    L^2(r_{\rm p}) = (\rg c)^2 \frac{2 (r_{\rm p}/\rg)^2}{r_{\rm p}/\rg - 2},
\end{equation}
which reduces to $[L(r_{\rm p})/(\rg c)]^{2} = 2~r_{\rm p}/\rg$ in the Newtonian limit. Therefore, the cross section for full disruptions is $\propto L^{2}(\physrad) \equiv \mathcal{L}_{\rm t}^2$, while the Newtonian angular momentum associated with a pericenter of $\rtidal$ we call $L_{\rm N}^2 = 2 G M_{\rm BH}r_{\rm t}$.

To demonstrate how our calculations for full disruption cross sections compare to estimates made without regard to either internal stellar structure or relativity, we examine in Figure~\ref{fig:com_NT_GR} the ratio of our calculation of the total disruption cross section to the estimate neglecting relativity and stellar structure, i.e., $(\Lphysr^2 - L_{\rm dc}^2)/(L_{\rm N}^2-L_{\rm N,dc}^{2})$ for $M_{\star}=0.3$, $1$ and $3$. Here, $L_{\rm dc} = 4r_{\rm g} c$ 
is the angular momentum of a parabolic direct capture orbit, and $L_{\rm N,dc}=2\rg c$. 

Relativistic effects become important only for $M_{\rm BH} \gtrsim 10^7$. When they do enter significantly, the cross section for total disruptions of low-mass stars rises sharply relative to the simple prediction, reaching a factor $5\times$ greater for $M_{\rm BH}=3 \times 10^7$. For these stars, the greater strength of relativistic tidal stress is the dominant mechanism.  However, for high-mass stars, the dominant effect is the growing importance of direct capture, and, relative to the simple estimate, the cross section for total disruptions falls.

\section{Energy distribution and characteristic energy width of stellar debris for full disruptions}\label{sec:energyscale}

In the conventional picture \citep{Rees1988}, the distribution function of debris mass with energy $dM/dE$ is approximated as flat from $-\Delta\epsilon$ to $+\Delta\epsilon$, and vanishes outside that range. The magnitude of $\Delta\epsilon= {GM_{\rm BH}R_{\star}}/{r_{\rm t}^{2}}$ was estimated by \citet{Rees1988} on the basis of tidal spin-up; the same estimate results from consideration of the spread of potential energy within the star when its distance from the black hole is some fiducial value \citep{Lacy+1982,Lodato+2009,Stone+2013,ServinKesden2017}; this fiducial value is chosen variously to be $r_{\rm p}$ \citep{Lodato+2009} or 
$\rtidal$ \citep{Stone+2013}. We use the latter in the definition of $\Delta\epsilon$.

To determine the orbital energy of the debris from our simulation data, we transform the 4-velocity at each boundary cell from the simulation box coordinates following the star frame to the Schwarzchild coordinates associated with the black hole. The energy $E$ is then given by $-1 - u_{\rm t}$ evaluated in the Schwarzchild frame (we use a $-+++$ signature). Figure~\ref{fig:dmde} shows the energy distribution $dM/dE$ for three different stellar masses (see Figure 8 in \citetalias{Ryu2+2019} for $dM/dE$ for five more values of $M_{\star}$). Although not far from flat, particularly for low-mass stars, it is not exactly flat, and in units of $\Delta\epsilon$ tends to be wider for higher $M_{\star}$. Nonetheless, it still retains a sharp edge, clearly defining a characteristic value of the energy.
We define this characteristic width of the distribution function $\Delta E$ such that the mass within $|E|<\Delta E$ is 90\% of the total debris mass.

The dependence of $\Delta E/\Delta \epsilon$ upon both $M_{\star}$ and $M_{\rm BH}$ can be reasonably well-described by an analytic form related to the ones we have proposed for $\physrad/\rtidal (\equiv\Psi)$:
\begin{equation}\label{eq:energy_fit}
\Xi(M_{\star},M_{\rm BH})\equiv\frac{\Delta E}{\Delta\epsilon}\simeq \Psi_{\rm BH}^{-1}\left[\frac{2}{1 + \Psi_{\star}^2}\right] .
\end{equation}
As shown in Figure~\ref{fig:dmde_maxrange}, this expression does a fairly good job matching the trends with $M_{\rm BH}$ (the \textit{left} panel) and $M_{\star}$ (the \textit{right} panel) found in our data.  Its form also implies that, like $\physrad/\rtidal$, $M_{\rm BH}$-dependence (expressed by $\Psi_{\rm BH}^{-1}$) and $M_{\star}$-dependence (the factor $2[1+\Psi_{\star}^{2}]^{-1}$) are separable.

\begin{figure*}
	\centering
		\includegraphics[width=8.9cm]{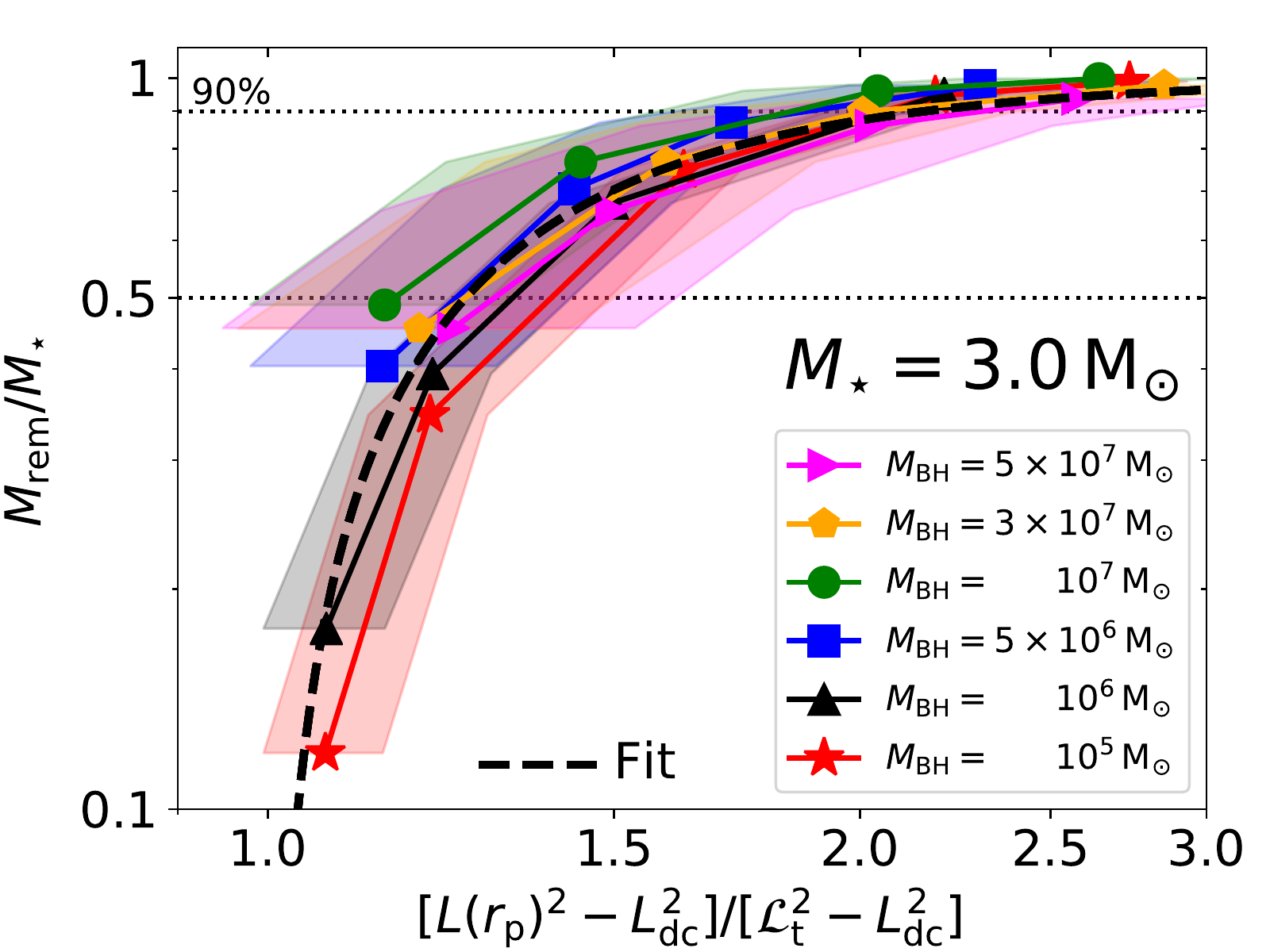}	
	\includegraphics[width=8.9cm]{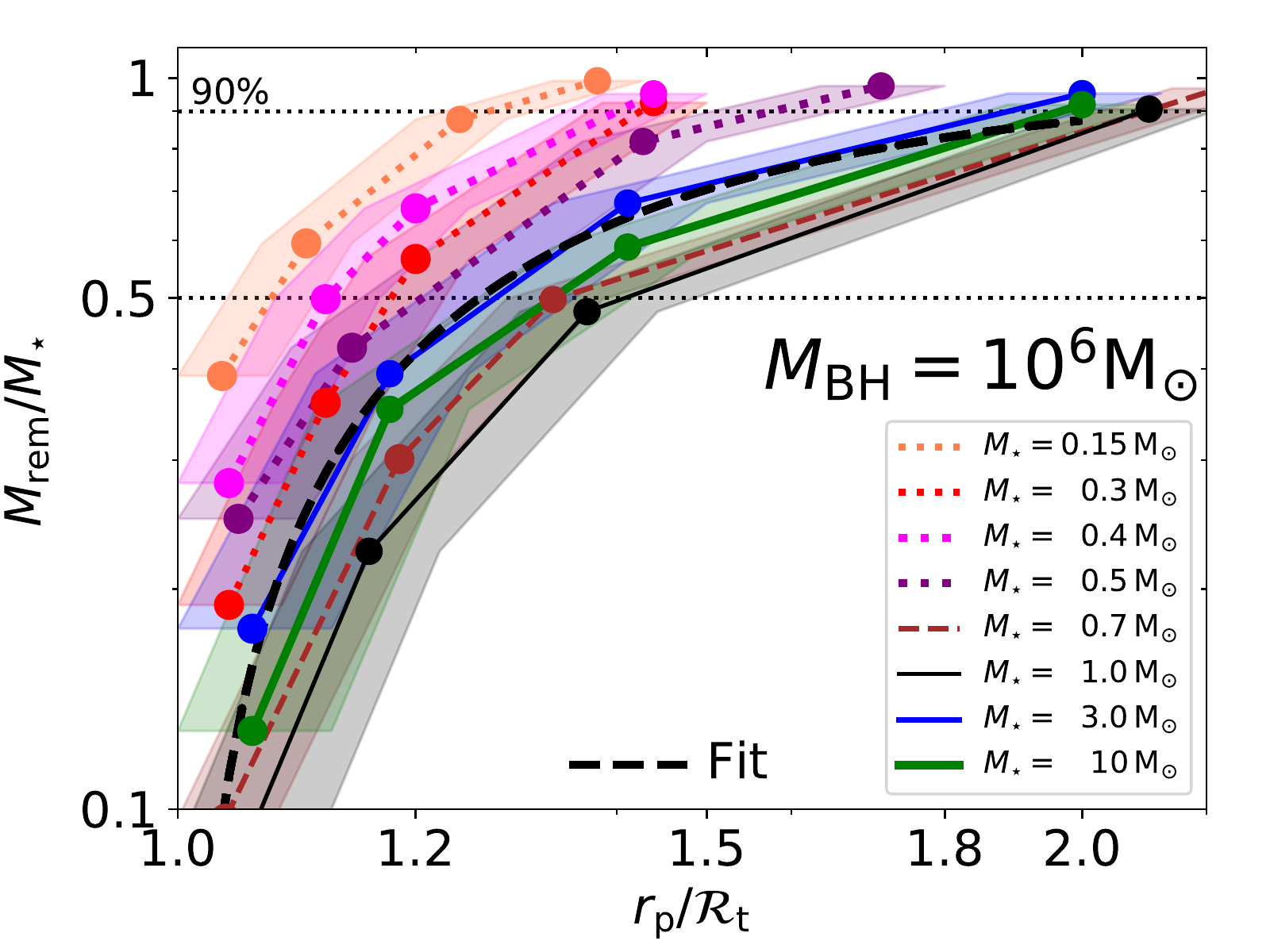}
	\caption{The fractional remnant mass $M_{\rm rem}/M_{\star}$. (\textit{Left} panel) $M_{\rm rem}/M_{\star}$ for $M_{\star}=3$ stars disrupted by BHs with various $M_{\rm BH}$ as a function of the ratio of the cross-section for all disruptions to the cross section for full disruptions.  (\textit{Right} panel) $M_{\rm rem}/M_{\star}$ as a function of $r_{\rm p}/\physrad$ for stars of eight different masses disrupted by a $10^{6}\Msol$ BH. In both panels, the shaded regions around the solid lines demarcate the ranges determined by the uncertainties of $\mathcal{R}_{\rm t}$, filled with the same colors as the solid lines. The uncertainty in $\mathcal{R}_{\rm t}$ is due to our discrete sampling of $r_{\rm p}$ ($0.05-0.1$ in $r_{\rm p}/r_{\rm t}$). The dotted horizontal lines show the 50\% and 90\% remnant mass-fraction levels. The fitting formula is plotted using a black dashed line in each panel, Equation~\ref{eq:remnant_mass_L} (\textit{left} panel) and Equation~\ref{eq:remnant_mass_rp} (\textit{right} panel).  }
	\label{fig:remnant_mass}
\end{figure*}

 Equation~\ref{eq:energy_fit} also has a significant implication: the two key quantities defining the properties of TDEs, the physical tidal radius and the characteristic energy width, are related. However,
 they are not related in a trivial way with $\Delta E \propto \physrad^{-1}$. 
 If one wishes to rewrite Equation~\ref{eq:energy_fit} in terms of a single radius, that radius is a complicated function of $\rtidal$ and $\physrad$. 
 Nonetheless, it is in this way that the energy width, which sets the characteristic fallback time and therefore the peak fallback rate (Section~\ref{subsec:para_fallback}), can be found directly from $\rtidal$ supplemented by $\Psi_{\star}$ and $\Psi_{\rm BH}$.

The nature of the $M_{\star}$- and $M_{\rm BH}$-dependence of $\Delta E/\Delta\epsilon$ is that, at fixed $M_{\rm BH}$, $\Delta E$ is smaller in magnitude than $\Delta\epsilon$ for the low-mass, less centrally-concentrated stars, 
while it is larger in magnitude for the high-mass, more centrally-concentrated stars. 
At fixed $M_{\star}$, $\Delta E/\Delta \epsilon$ diminishes with increasing $M_{\rm BH}$.

If an even closer fit is required than the simple one defined in Equation~\ref{eq:energy_fit}, one can instead make use of two fitting formulae, of which the first substitutes for  $\Psi_{\rm BH}^{-1}$ in Equation~\ref{eq:energy_fit}, while the second replaces $2/(1+\Psi_{\star}^2)$:
\begin{align}\label{eq:Xi_BH}
\Xi_{\rm BH}(M_{\rm BH}) &=1.27 - 0.300\Mbh^{0.242},\\ \label{eq:Xi_star}
\Xi_{\star}(M_{\star}) &= \frac{0.620+\exp{[(M_{\star}-0.674)/0.212]}}{1 + 0.553~\exp{[(M_{\star}-0.674)/0.212]}}.
\end{align}
These expressions replicate our numerical results with very small errors (the dashed lines in both panels of Figure~\ref{fig:dmde_maxrange}).

\section{Remnant mass in partial disruptions}\label{sec:partial}

Stars are partially disrupted and lose some fraction of their masses when $\physrad<r_{\rm p} < \widehat{R}_{\rm t}$. $\widehat{R}_{\rm t}$ refers to the largest pericenter distance yielding partial disruptions, which is a few times $\physrad$ (\citetalias{Ryu3+2019}, \citetalias{Ryu4+2019}).   Substantial mass-loss (e.g., as defined by $M_{\rm rem}/M_{\star} < 0.7$) occurs for predominantly radiative stars for $r_{\rm p}/\physrad \lesssim 1.5$, but for a somewhat smaller range of pericenters for convective stars.

Remarkably, the most noteworthy parameter characterizing a partial disruption, the mass retained by the star at the end of the event, can be simply related to the star's angular momentum $L$, making it easy to estimate the rate at which partial disruptions of a given character occur.  As illustrated in the \textit{left} panel of Figure~\ref{fig:remnant_mass}, we find that for partial tidal disruptions by a non-spinning black hole it is a good approximation---independent of both $M_{\star}$ and $M_{\rm BH}$---to write
\begin{equation}\label{eq:remnant_mass_L}
\frac{M_{\rm rem}}{M_{\star}} = 1 - \left[\frac{L(r_{\rm p})^{2}-L_{\rm dc}^{2}}{\mathcal{L}_{\rm t}^{2}-L_{\rm dc}^{2}}\right]^{-3}.
\end{equation}
In the Newtonian limit, this expression reduces to 
\begin{align}\label{eq:remnant_mass_rp}
    \frac{M_{\rm rem}}{M_{\star}} = 1 - \left(\frac{r_{\rm p}}{\physrad}\right)^{-3}.
\end{align}
which is depicted in the \textit{right} panel of Figure~\ref{fig:remnant_mass}, showing the behavior of different mass stars when $M_{\rm BH}$ is small enough that the Newtonian limit does not create a large error.  The full expression (Equation \ref{eq:remnant_mass_L}) is surprisingly good considering the wide range of $M_{\star}$ and $M_{\rm BH}$ to which it can be applied.  For $M_{\star}=3$, the curve defined by Equation ~\ref{eq:remnant_mass_L} runs right down the middle of the $M_{\rm rem}/M_{\star}$ curves for specific black hole masses. The largest departure (for the smallest $M_{\rm rem}/M_{\star}$ with $M_{\rm BH}=10^{5}$) is a factor $\simeq 1.5$; the second largest is only a factor $\simeq 0.7$, and the rest are small enough that the mean fractional error is $\simeq 10\%$.  For other high-mass stars, the fit is almost as close.  On the other hand, for lower stellar masses, particularly $M_{\star} < 0.4$, the fit performance is poorer (the \textit{right} panel of Figure~\ref{fig:remnant_mass}); for $M_{\star}=0.15$, the fit's functional dependence of $M_{\rm rem}/M_{\star}$ on $r_{\rm p}/\physrad$ is correct, but its magnitude is too small by a factor $\sim 4$.

\section{Implications}
\label{sec:discussion}

$M_{\star}$-dependence, manifested through internal structure, and $M_{\rm BH}$-dependence, due to relativistic effects, lead to significant changes in observable quantities. Changes in the range of pericenters producing tidal disruptions translate directly into changes in event cross sections. Because the debris energy distribution determines the debris orbital period distribution, these changes alter the predicted fallback rate. 

Our analytic fits to $\Psi$ and $\Xi$ enable us to transform the simple conventional formulae linking stellar mass and black hole mass to mass fallback properties into more accurate expressions.  The quantitative contrast with the older formalism can have significant implications for observations; we consider a few here.

\subsection{Tidal disruption rates}
\label{sub:r_tderate}

\subsubsection{The ``full loss-cone" regime}

The rate of full TDEs depends on $\physrad$ or, equivalently, $L(\physrad) \equiv \mathcal{L}_{\rm t}$ (Equation~\ref{eq:L}). However, the nature of this dependence varies with circumstances \citep{FrankRees1976,LightmanShapiro1977,Alexander2005,Merritt2013}.  When the angular momentum distribution of stellar orbits is smooth across all directions, including those implying passage very close to the nuclear black hole, it is appropriate to use the cross section formalism, in which the rate of events with pericenters $\leq \physrad$ is $\propto \Lphysr^{2}$.

This is the case, for example, when a number of possible mechanisms operate, e.g., resonant relaxation \citep{RauchTremaine1996,RauchIngalls1998} or triaxiality in the stellar cluster \citep{Merrittbook2013}.
On the other hand, if the only mechanism influencing the stellar angular momentum distribution is gravitational encounters with individual stars, the situation changes in a way that depends upon the magnitude of the ratio between $(\Delta L)^2$, the per-orbit mean-square change in angular momentum, and $\Lphysr^2$.
The stellar distribution function remains smooth in all directions only when $\Delta L^2/\Lphysr^2 > 1$ (the ``full loss-cone" or ``pinhole" regime).

Whenever it is appropriate to speak in terms of cross sections, our results translate directly into alterations to the event rates. If the  total disruption cross section were $\propto L^2(\rtidal)$, the  stellar mass-dependence of the rate would be $\propto M_{\star}^{-1/3}R_{\star}\propto M_{\star}^{0.55}$ (see \citetalias{Ryu2+2019}). However,  we find that $\Lphysr^2$ is nearly independent of $M_{\star}$ for all $M_{\star}\lesssim 3$.  For all $M_{\rm BH} < 10^7$, the value of this cross section is $\simeq 1/2 \times$ the value derived from $\rtidal$ for $M_{\star}=1$, after allowance for direct captures, whose cross section is $\propto L_{\rm dc}^2$. \citep{Kesden2012}.
At higher black hole masses, the cross section for total disruptions relative to the Newtonian prediction can either increase or decrease, depending on stellar mass, as two relativistic effects, greater tidal stress and direct capture, compete (Figure~\ref{fig:com_NT_GR}).

However, the near-independence of $\physrad$ with respect to $M_{\star}$ combined with the decoupling between $\Psi_{\star}$ and $\Psi_{\rm BH}$ means that the absolute cross section for tidal disruption remains nearly independent of $M_{\star}$ for all black hole masses.  This implies that the suppression of total tidal disruptions by competition with direct capture is {\it nearly independent} of $M_{\star}$.  For example, in Schwarzschild spacetime, more than 50\% of all events with pericenter $\lesssim 14~r_{\rm g}$ produce direct captures; this corresponds to a factor of $>2$ suppression of the complete tidal disruption rate when $M_{\rm BH} \gtrsim 5 \times 10^6$. The suppression rises to a factor $>10$ when $M_{\rm BH} \gtrsim 3 \times 10^7$ (See Figure 7 in \citetalias{Ryu4+2019}).

\subsubsection{The ``empty loss-cone" regime}

On the other hand, when $ (\Delta L)^2 < \Lphysr^2$, the stellar distribution function is said to evolve in the ``empty loss-cone" or ``diffusive" regime.  It has been generally thought that in these circumstances, most stars approach the edge of the loss-cone only gradually, taking small steps up and down in $L^2$ \citep{LightmanShapiro1977,Merritt2013}, while occasionally being scattered more strongly \citep{Weissbein2017}.  This regime is often associated with lower orbital energy \citep{Alexander2005,StoneMetzger2016}. Once a star slips inside the loss-cone, it is destroyed at its first pericenter.   Consequently, the majority of the stars within the loss-cone have velocities directed very close to its edge, making the ``cross section" language inappropriate because the distribution of impact parameters is not uniform. The event rate then depends only logarithmically on $\Lphysr$ \citep{LightmanShapiro1977,Merritt2013,Weissbein2017}.

However, while stars move through the angular momentum range just outside the loss-cone, i.e., $\Lphysr^2 < L^2 \lesssim 3\Lphysr^2$ (see Figure~\ref{fig:remnant_mass}), they suffer partial disruption every time they pass through pericenter.  Consequently, normal full disruptions do not happen in this regime.  Instead, several alternative pathways to destruction may be followed.

One possibility is that, as discussed at greater length in \citetalias{Ryu3+2019}, some remnants created by partial disruptions may not relax to main sequence structure in a single orbit. Remnants when first formed are considerably more distended than a main sequence star of that mass because they are hotter and rotate more rapidly.  If their  relaxation time  (in particular for thermal properties) is longer than an orbit, the value of $\Lphysr$ corresponding to their distended structure would be rather larger than it would be for a main sequence star of mass $M_{\rm rem}$.  Because the specific angular momentum of the remnant is essentially unchanged, its pericenter would then be well inside its new $\physrad$, and it would be fully disrupted one orbit later.  The full disruption of such a perturbed remnant would be quite different from a full disruption of an ordinary main sequence star because the quantitative characteristics of tidal disruptions depend significantly on the star's internal structure.

Another possible outcome is that the orbital energy of a remnant could be sufficiently larger than the initial value that $(\Delta L)^2$ on the new orbit becomes large relative to $\Lphysr^2$.  From this point on, such a star would evolve in the full loss-cone regime. This process can also act in the opposite direction: a star may enter a partial disruption event with large $(\Delta L)^2/\Lphysr^2$ and exit it as a remnant whose specific energy is small enough to make $(\Delta L)^2/\Lphysr^2$ small.  In this manner, partial disruptions can act as a transfer channel between scattering regimes.
 
There is also a third way stars could evolve if the remnants' orbital energies are such as to keep them in the regime of small angular momentum change per orbit, and they relax to main sequence structure within an orbit.  In this case, their pericenters remain in the range associated with partial disruptions. Consequently, their mass decreases at each pericenter passage \citep{StoneMetzger2016}.  Although the value of $\Lphysr$ changes as each partial disruption diminishes the star's mass, it does not change much; as we have shown in Section~\ref{sec:physicalR}, $\physrad$ depends only weakly on $M_{\star}$ for main sequence stars with $M_{\star} \lesssim 3$.   The relation between $M_{\rm rem}/M_{\star}$ and $L^2$ we have uncovered shows that this mass-loss can be substantial: once $L^2 \lesssim 1.4\Lphysr^2$, $M_{\rm rem}/M_{\star} < 1/2$.  Moreover, because their progress through the partial disruption zone is stochastic, these stars are likely to suffer numerous partial disruptions before their angular momentum becomes small enough that they can be totally disrupted: if $(\Delta L)^2/\Lphysr^2$ is independent of $L$, $\simeq [(\Delta L)^2/\Lphysr^2]^{-1}$ orbital periods are required to wander from $\simeq 3\Lphysr^2$ to $\Lphysr^2$.  Thus, their mass upon total disruption will, in most cases, be a small fraction of their initial mass.  Only two effects limit this process. One, which has probability $\sim 10\%$, is that this chain of partial disruptions is interrupted by a strong scattering event, and the star either goes into the loss-cone or far outside the zone of partial disruptions \citep{Weissbein2017}.  The other is the trend for stars with $M_{\star} \lesssim 0.4$ to retain a larger fraction of their mass than more massive stars do (see the \textit{right} panel of Figure~\ref{fig:remnant_mass}) . The overall result of this third possibility would be to reduce sharply the rate of total disruptions of massive stars, while only slightly enlarging that of the least massive stars compared to that of high-mass stars because low-mass stars dominate the general stellar population.

Each of these scenarios has a different effect on the event rate. When the first  applies, it effectively changes the rate of total disruptions from the rate at which stars cross the loss-cone boundary by diffusion to the rate given by the cross section for a partial disruption severe enough to materially disturb the structure of the remnant created.  It also entails one substantial partial disruption for each (non-standard) total disruption.  The second effectively makes the angular momentum range associated with partial disruptions a venue for substantial change in the stars' specific energy.   The third does not alter the event rate predicted in the small $(\Delta L)^2/\Lphysr^2$ regime, but it materially changes the distribution in stellar mass for total disruptions.  It is also associated with a large number of partial disruptions, of varying severity, for each eventual total disruption of a star that has gone through this process.

\subsection{Orbital properties of stellar debris}\label{subsub:debris}

\subsubsection{Testing the ``frozen-in" approximation}

An heuristic argument dubbed ``the frozen-in approximation" has sometimes been used as a device to estimate the characteristic energy width of the debris $\Delta E$ in terms of the spread in potential energy across the star near the beginning of the encounter \citep{Lodato+2009,Stone+2013}.  The basic idea is that from some special moment onward, individual fluid elements within the star travel on ballistic orbits.   When the point at which the energy spread is evaluated is independent of the actual orbital pericenter (e.g., $\rtidal$ or $\physrad$), $\Delta E$ is the same for all $r_{\rm p} \leq \physrad$.

As we have shown in Section~\ref{sec:energyscale}, applying the ``frozen-in approximation" at any of the plausible choices for the fiducial point ($\rtidal$, $r_{\rm p}$, or $\physrad$) always leads to errors in $\Delta E$ at the factor $\sim 2$ level.  This fact suggests that the debris energy is influenced by more than just the matter's initial location in the star and the instantaneous tidal potential at some special location.  Hydrodynamics, stellar self-gravity and the nature of the tidal stress over an extended range of distances from the BH are also important.  In fact, in \citetalias{Ryu2+2019}, we show that stars undergoing tidal encounters lose mass over a wide range of orbital separations from the BH, from near $r_{\rm p}$ to a distance an order of magnitude larger (\citealt{Guillochon+2013} also reported a version of this behavior), a finding that also points toward no single location having a unique role in determining the debris energy distribution.

Moreover, the fluid elements of the star do not follow ballistic orbits as they pass near the black hole and then travel outward until finally shed.  During the entire period of mass-loss, the star's instantaneous distance from the black hole is quite close to its ``instantaneous tidal radius", the one computed on the basis of the star's density and distance from the black hole as functions of time (see \citetalias{Ryu2+2019} for a lengthier discussion of the instantaneous tidal radius). Consequently, for the extended period of time while the star is torn apart, stellar self-gravity and black hole tidal gravity are comparable in magnitude (as are pressure forces as well).   Similar behavior was seen in the simulations of \citet{Steinberg+2019}, who studied non-relativistic disruption dynamics for $r_{\rm p}/\rtidal \simeq 0.14 - 0.2$.  All these results are contrary to the motivation of the frozen-in approximation, which \citet{Lodato+2009} explicitly describe as an ``impulse" approximation.  Note, however, that this conclusion does not remove the possibility that $\Delta E$ is independent of $r_{\rm p}$ for $r_{\rm p} \leq \physrad$ \citep{Stone+2013}.  Our simulation data are consistent with this possibility, but do not span a wide enough range in $r_{\rm p}$ to test it credibly.  It is possible that some mechanism other than a freezing of fluid element energy enforces this outcome.

\subsubsection{Peak mass return: time and rate}\label{subsec:para_fallback}

The peak return rate and the time at which this peak is reached are quantities of particular interest to predictions of a TDE's light output. The two corrections introduced here, for stellar internal structure and for general relativistic effects, both alter these parameters' dependence on $M_{\star}$ and $M_{\rm BH}$.  These changes are encapsulated in the change to the debris energy, $\Xi \equiv \Delta E/\Delta\epsilon$.  However, as we shall see, the stellar internal structure also changes the fallback rate in a subtler way.

The mass fallback rate of stellar debris on ballistic orbits is \citep{Rees1988,Phinney1989} 
\begin{align}
\dot{M}_{\rm fb}=\frac{dM}{dE}\left|\frac{dE}{dt}\right|=\frac{(2\uppi G M_{\rm BH})^{2/3}}{3}\frac{dM}{dE}t^{-5/3}.
\label{eq:mdot}
\end{align}
The steepness of the rise to $\dot M_{\rm peak}$ (peak mass return rate) and the specific shape of that peak depend on the shape of $dM/dE$ for the most tightly-bound matter.  When $dM/dE$ has a sharp, steep edge at $E = - \Delta E$, as it does for low-mass stars (\citetalias{Ryu2+2019}), independent of $M_{\rm BH}$ (\citetalias{Ryu4+2019}), the rise to the peak is also comparatively sharp.  On the other hand, when there is a noticeable wing extending to energies below $-\Delta E$, as is the case for high-mass stars disrupted by low-mass black holes, the rise is more gradual (\citetalias{Ryu2+2019}, \citetalias{Ryu4+2019}).  It is the impact of these effects on the time-dependence of mass fallback that we encapsulate in the factor $f$ in Equation~\ref{eq:peak_mdot}.

Because of the differing shapes of the fallback rate curves, we define $t_{\rm peak}$ as the time at which 5\% of $M_{\star}$ has returned to the black hole. This time corresponds to the time of the absolute maximum when the peak is sharp (for all stars being disrupted by high-mass black holes, and low-mass stars for black holes of any mass), and the beginning of the maximum when the peak is relatively flat (for high-mass stars encountering low-mass black holes).

\begin{figure}
	\centering
	\includegraphics[width=8.6cm]{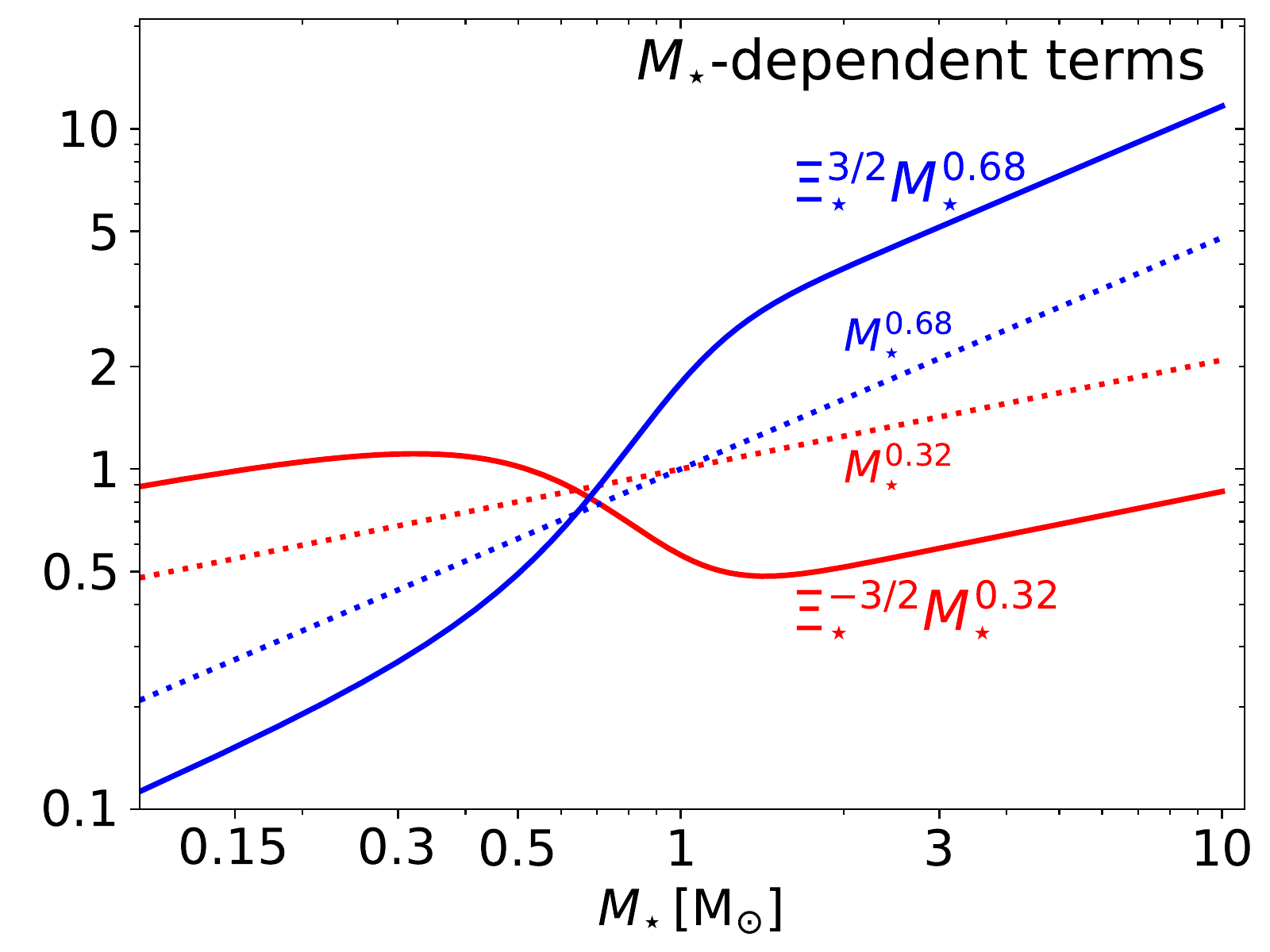}
	\includegraphics[width=8.6cm]{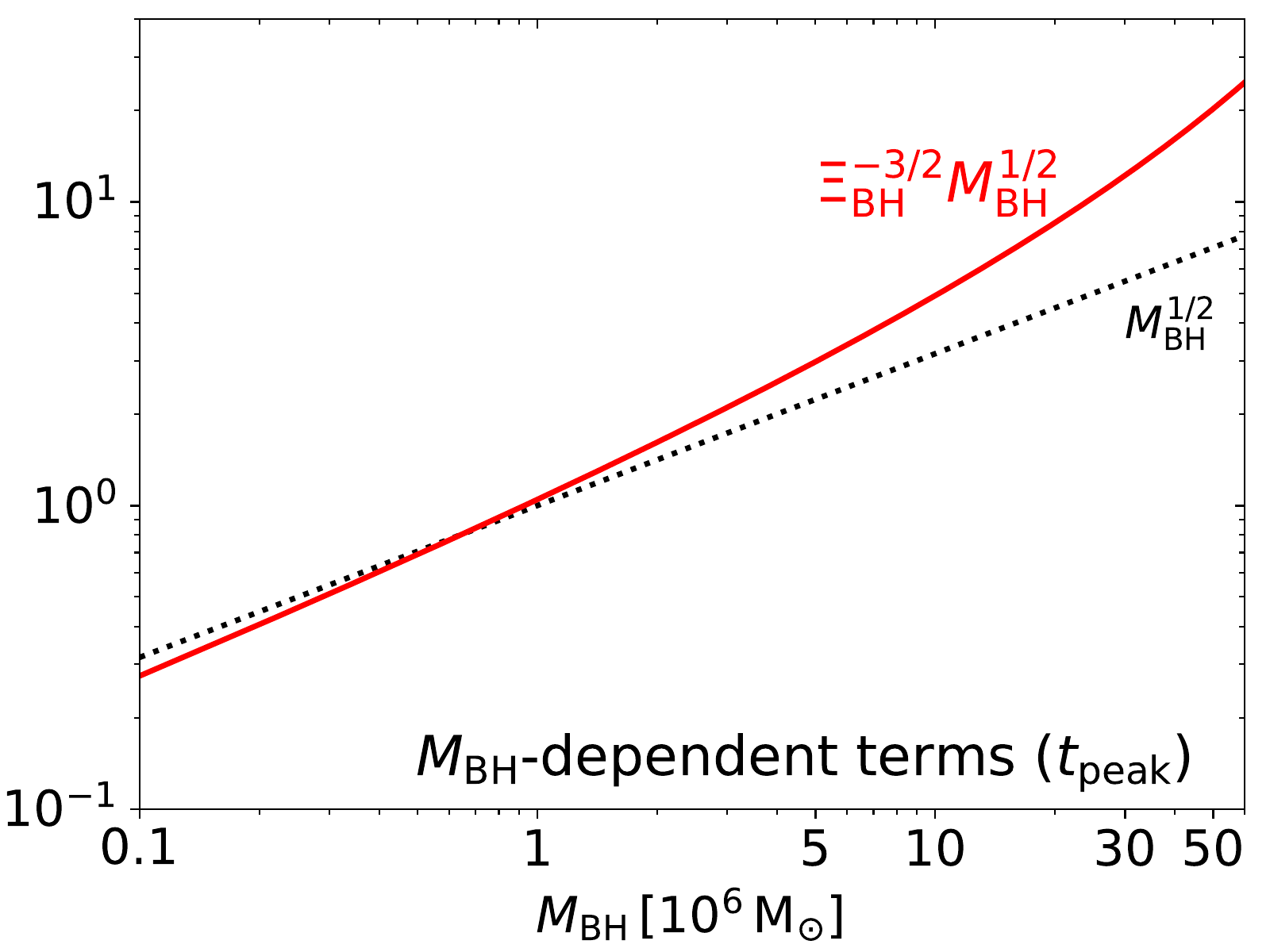}	
	\includegraphics[width=8.6cm]{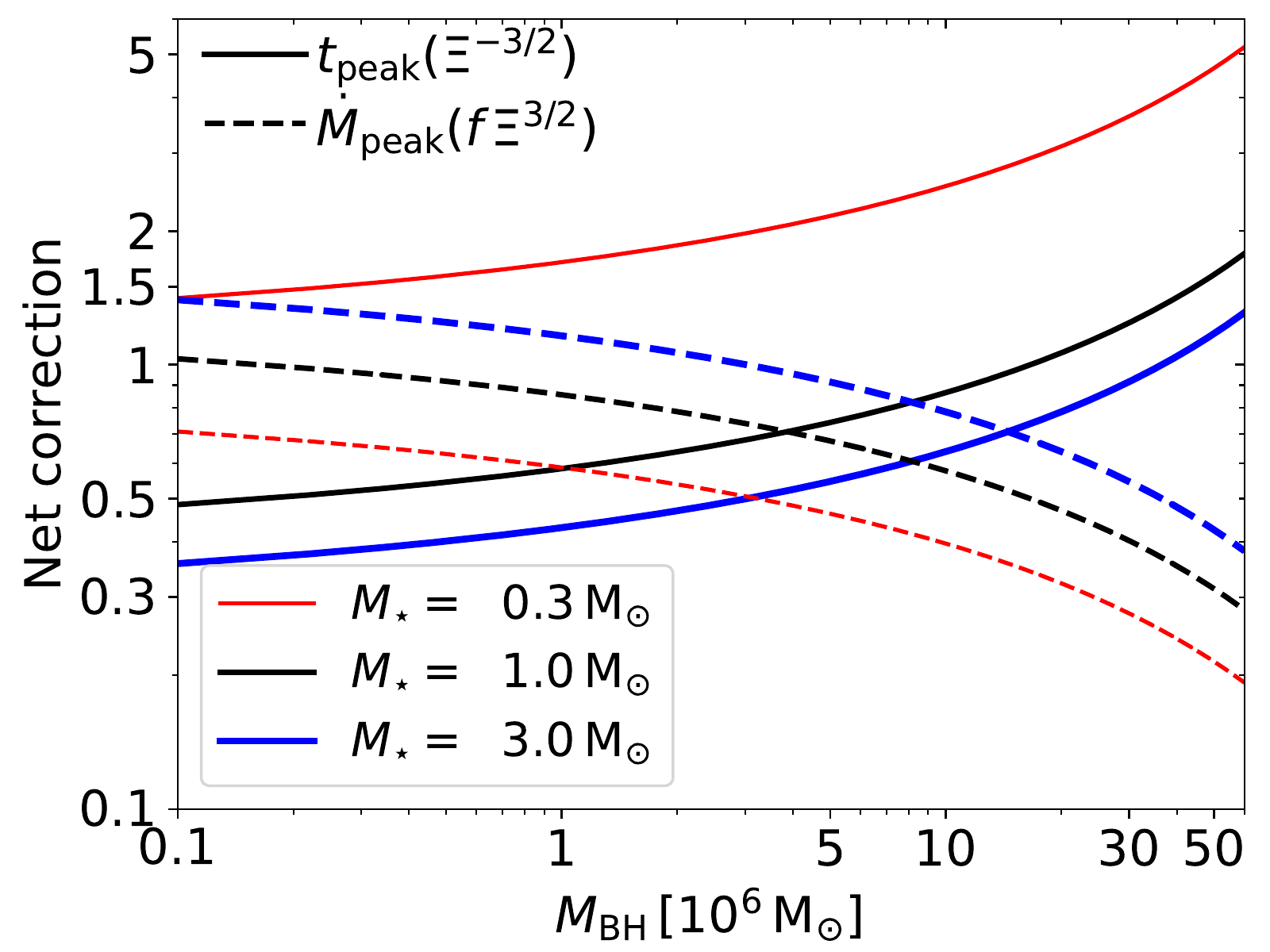}	
	\caption{(\textit{Top}) $M_{\star}$-dependence of $t_{\rm peak}$ (red) and $\dot{M}_{\rm peak}$ (blue).  Solid curves use our most precise fit to $\Xi_{\star}$ (Equation~\ref{eq:Xi_star}); dotted use the traditional prediction. (\textit{Middle}) $M_{\rm BH}$-dependence of $t_{\rm peak}$, using our fit (Equation~\ref{eq:Xi_BH}) to $\Xi_{\rm BH}$ (red solid) and the traditional form (black dots).  The $M_{\rm BH}$-dependence of $\dot{M}_{\rm peak}$ is the inverse of that of $t_{\rm peak}$. (\textit{Bottom}) Net correction factor for three stellar masses: $0.3\Msol$ (red), $1\Msol$ (black), and $3 \Msol$ (blue); for $t_{\rm peak}$ (solid lines) and for $\dot{M}_{\rm peak}$, including the $f$ correction (Equation~\ref{eq:peak_mdot}, dashed lines).}
	\label{fig:correction_factor}
\end{figure}

In terms of  $\Xi$, the peak fallback time $t_{\rm peak}$ is 
\begin{align}
\label{eq:peak_t}
t_{\rm peak}&=\frac{\uppi}{\sqrt{2}}
\frac{G M_{\rm BH}}
{\Delta {E}^{3/2}},\nonumber\\
&\simeq0.11\yr~\Xi^{-3/2} M_{\star}^{-1}R_{\star}^{3/2}\Mbh^{1/2},
\end{align}
and the peak fallback rate $\dot{M}_{\rm peak}$ at $t=t_{\rm peak}$ is
\begin{align}
\label{eq:peak_mdot}
\dot{M}_{\rm peak}&\simeq f \frac{M_{\star}}{3t_{\rm peak}},\nonumber\\
&\simeq1.49\Msol \yr^{-1}\left(\frac{f}{0.5}\right)
\Xi^{3/2}M_{\star}^{2}R_{\star}^{-3/2} \Mbh^{-1/2},
\end{align}
where the correction factor $f$ accounts for the different shape of the energy distribution near the tails (see discussion below). We take $f=1$ for $M_{\star}\leq0.5$ and $0.5$ for $M_{\star}>0.5$. 

To isolate the dependence of $t_{\rm peak}$ and $\dot{M}_{\rm peak}$ on $M_{\rm BH}$ and $M_{\star}$, we first split $\Xi$ into its portions dependent on $M_{\star}$ and $M_{\rm BH}$. Next we insert the $M_{\star}-R_{\star}$ relation for our main sequence stellar models, i.e., $R_{\star}\propto M_{\star}^{0.88}$ (\citetalias{Ryu2+2019}). Equations \ref{eq:peak_t} and \ref{eq:peak_mdot} then become
\begin{align}
\label{eq:peak_t2}
t_{\rm peak}&\propto \left[\Xi_{\star}^{-3/2} \Mstar^{0.32}\right]\left[ \Xi_{\rm BH}^{-3/2} M_{\rm BH}^{1/2}\right],\\
\label{eq:peak_mdot2}
\dot{M}_{\rm peak}&\propto \left[\Xi_{\star}^{3/2} \Mstar^{0.68}\right] ~\left[\Xi_{\rm BH}^{3/2} M_{\rm BH}^{-1/2}\right].
\end{align}
The \textit{top} panel of Figure~\ref{fig:correction_factor} displays the effect on the $M_{\star}$-dependence due to our corrections.  Compared to the simple approximation in which $\Delta E = \Delta \epsilon$, $t_{\rm peak}$ is later, by a little less than a factor of two, for low-mass stars and sooner, by a little more than factor of two, for high-mass stars.  Only for $M_{\star} \simeq 0.7$ do $t_{\rm peak}$ and $\dot M_{\rm peak}$ coincide with the prediction of the simple approximation.  The scaling of both $t_{\rm peak}$ and ${\dot M}_{\rm peak}$ with $M_{\star}$ is close to the approximate scaling for both ends of the stellar mass spectrum, but differs strongly from $M_{\star} \simeq 0.5$ to $M_{\star} \simeq 1.5$.  Overall, realistic internal structures cause $t_{\rm peak}$ to have only a weak net trend as a function of $M_{\star}$ and to make this dependence non-monotonic, in contrast with the simple estimate's prediction that $t_{\rm peak} \propto M_{\star}^{0.32}$.  For $\dot M_{\rm peak}$, realistic internal structures {\it steepen} the global $M_{\star}$-dependence, with the sharpest dependence concentrated in the range $0.5 \lesssim M_{\star} \lesssim 1$.

The \textit{middle} panel of Figure~\ref{fig:correction_factor} shows how our corrections alter $M_{\rm BH}$-dependence.  There is little departure from the traditional dependence until $M_{\rm BH} > 5 \times 10^{6}$,  but for more massive black holes, $t_{\rm peak}$ occurs later, by as much as a factor of 3 for $M_{\rm BH} = 5 \times 10^7$. Because the relativistic correction factor for $t_{\rm peak}$ (i.e., $\Xi_{\rm BH}^{-3/2}$) varies with $M_{\rm BH}$ in the same sense that the Newtonian expression does, these effects result in stronger dependences of $t_{\rm peak}$ and $\dot{M}_{\rm peak}$ on $M_{\rm BH}$ than are predicted by Newtonian dynamics. Moreover, relativistic corrections become dominant over the Newtonian $M_{\rm BH}$-dependence ($d\ln \Xi_{\rm BH}^{-3/2}/d\ln M_{\rm BH}>0.5$) for $M_{\rm BH}>3\times10^{7}$.

These corrections can have a significant impact on parameter inference resting on mass-return rates.  For example, as shown in the \textit{upper} panel of Figure~\ref{fig:correction_factor}, neglecting the internal structure correction factor for $M_{\star}$ could lead to inferring $M_{\star}=0.3$ from a measurement of $t_{\rm peak}$ when the real $M_{\star} \simeq 0.8$. Similarly, when Newtonian analysis of $t_{\rm peak}$ would indicate $M_{\rm BH} = 10^7$, a properly relativistic approach leads to $M_{\rm BH} \simeq 5 \times 10^6$. Therefore, it is imperative to take into account both corrections (the \textit{bottom} panel of Figure~\ref{fig:correction_factor}) for more accurate inferences of $M_{\star}$ and $M_{\rm BH}$.  Analyses neglecting both of these effects  \citep[e.g.,][]{Mockler2019}
incur systematic errors of factors of several in both the inferred stellar mass and the inferred black hole mass, on top of whatever additional systematic errors might be present due to other aspects of the lightcurve modeling (e.g., direct identification of light output with mass return rate).

\subsubsection{Unbound debris energy and speed at infinity}
\label{dis:unboundebris}

The energy of the most highly-bound matter determines the time of peak mass-return; the energy of the most highly-{\it unbound} matter determines the fastest speed of the ejecta that never return to the black hole, as well as the total amount of energy available for deposition in surrounding gas. We find that the total energy of unbound debris is $\simeq4\times10^{50}~\Xi~ M_{\star}^{0.79}~(M_{\rm BH}/10^{6})^{1/3}\erg$ (Equations \ref{eq:deltae}, \ref{eq:Xi_star} and \ref{eq:Xi_BH}), and the greatest speed at infinity for the bulk of the ejecta mass is
$\simeq6\times10^{3}~\Xi^{1/2}~M_{\star}^{-0.11}~(M_{\rm BH} / 10^{6})^{1/6}\km\s^{-1}$; these scalings take into account the main sequence mass-radius relation (\citetalias{Ryu2+2019}).  Because $\Xi^{1/2}$ changes by at most a factor of 1.6 from low-mass stars to high-mass and a factor of 1.4 from the Newtonian limit to $M_{\rm BH} = 5 \times 10^7$, this speed is exceedingly weakly dependent on both $M_{\star}$ and $M_{\rm BH}$.

Both the energy and the mass of the unbound ejecta are comparable to those of a supernova remnant.  One might therefore expect that when the unbound debris shocks against whatever gas surrounds the black hole, there would be radio emission \citep{Guillochon+2016}.  Such emission has, in fact, been seen in several cases, with particularly rich datasets obtained from ASASSN-14li and CNSS J0019+00 \citep{Alexander+2016,vanVelzen+2016,Anderson+2019}. 
Using the equipartition formalism for relativistic synchrotron self-absorbed spectra  \citep{BarniolDuran+2013}, \citet{Krolik+2016} found that the linear scale of the emission region in ASASSN-14li expanded at a constant speed $\simeq (1.45 - 2)\times10^{4}$~km~s$^{-1}$ (see also \citet{Alexander+2016} for a comparable estimate), while \citet{Anderson+2019} performed a very similar analysis on CNSS~J0019+00 and likewise found a constant speed $\approx 1.5 \times 10^4$~km~s$^{-1}$.  Because these speeds are close to those expected for the fastest-moving unbound ejecta, \citet{Krolik+2016} suggested that the unbound ejecta are, indeed, responsible.

Our results strengthen that conclusion for two reasons.  First, the characteristic energy spread we find for $M_{\rm BH}=10^{6}$ is larger than the conventional estimate for all $M_{\star} > 0.7$, and larger by a factor $\simeq 1.8$ for $M_{\star} \gtrsim 3$. The ratio $\Xi$ of the characteristic energy width to the conventional estimate decreases as $M_{\rm BH}$ increases (see the {\textit{left}} panel of Figure~\ref{fig:dmde_maxrange}), but it nonetheless remains larger than unity over a wide range of $M_{\rm BH}$ and $M_{\star}$, e.g., for $M_{\star} \geq 1$ and $M_{\rm BH} \lesssim 10^{7}$. Second, the amount of mass required is only $\sim 10^{-4} - 10^{-2}\Msol$ (in ASASSN-14li, \citealt{Krolik+2016} inferred $\sim 10^{-4}\Msol$; in CNSS J0019+00, \citealt{Anderson+2019} found $7 \times 10^{-3}\Msol$).  This matter could therefore come from the power-law tail in $dM/dE$ that extends beyond $\Delta E$  when $M_{\star}>0.7\Msol$ (\citetalias{Ryu2+2019}).  For $M_{\rm BH} \sim 10^6$, there is $\sim 3 \times 10^{-4} M_{\star}$ at energies $\gtrsim 3\Delta\epsilon$ for all stars with $M_{\star} > 0.7$; although the wings narrow at higher black hole mass, for $M_{\rm BH} \sim 10^7$, the energy at this mass fraction is still $\simeq 2 \Delta\epsilon$ (\citetalias{Ryu4+2019}).  The speed of this material would then be $\geq 10,000 M_{\star}^{-0.11} \km\s^{-1}$ for black hole masses not too much more than $10^6$, or  $\geq 12500 M_{\star}^{-0.11}$~km~s$^{-1}$ for $M_{\rm BH} \sim 10^7$.
Alternatively, it may also be possible for a small amount of debris mass to reach such speeds in a highly-penetrating event \citep{Yalinewich+2019}.

\section{Summary}
\label{sec:summary}

In this paper we have presented the principal results from our suite of relativistic tidal disruption simulations using realistic main sequence stellar structures for stars of many different masses encountering black holes over a wide range of mass.  We have shown that $M_{\star}$-dependence in disruption properties (due to mass-dependent contrasts in internal structure) is largely decoupled from $M_{\rm BH}$-dependence (due to relativistic effects). 
The $M_{\star}$- and $M_{\rm BH}$-dependence of both $\physrad$ and $\Delta E$ due to these effects can be described quite accurately by analytic formulae.  Subsequent papers in this series will fill in the details, both of our methods and of our results; here we focused on our results' implications for observable properties such as event rates and the time-dependence of mass fallback.

Several broad themes can be seen clearly.  The order-of-magnitude estimate $r_{\rm t}$ for the maximum pericenter yielding a full disruption requires correction by factors $0.4-4$, depending on $M_{\star}$ and $M_{\rm BH}$.  One way to understand the $M_{\star}$-dependent part of this correction is a physically intuitive model that relates $\physrad$ to the {\it central} density of the star rather than to its mean density. Because the sense of the $M_{\star}$-dependent correction to $r_{\rm t}$ runs opposite to the dependence of stellar radius on stellar mass, the net result is a physical tidal radius that is roughly constant over the range $M_{\star}\simeq0.15-3$, a range spanning the overwhelming majority of all stars. On the other hand, for fixed $M_{\star}$, the ratio $\physrad/\rtidal$ increases with $M_{\rm BH}$. For $M_{\rm BH}\gtrsim 3\times10^{7}$, the location of the physical tidal radius is more sensitive to relativistic effects than to the Newtonian physics embodied in $r_{\rm t}$.  These corrections figure directly into rate predictions.

A comparable correction is required for the characteristic energy spread of the tidal debris.  The simple estimate is too large for low-mass stars and too small for high-mass stars when black hole masses are low. However, due to relativistic corrections, for higher black hole masses, the discrepancy between the simple estimate and the characteristic energy width becomes larger for low-mass stars, but smaller for high-mass stars.  Moreover, no version of the ``frozen-in" approximation can correctly predict this quantity, and total disruption of stars with $M_{\star} > 0.7$ generically produces power-law tails in the distribution function $dM/dE$ except for events involving very high-mass black holes. Alterations in the energy spread immediately imply differing fallback timescales, as the time of peak mass-return is $\propto (\Delta E)^{-3/2}$.  Neglect of these corrections when inferring $M_{\star}$ or $M_{\rm BH}$ from fallback histories could lead to significant errors.

The cross section for total disruptions of low-mass stars is larger than the simple estimate, while it is more than a factor of two smaller for high-mass stars.

Partial disruptions should occur at a rate comparable to total disruptions. The fraction of the star's mass lost in such an encounter can be described surprisingly accurately by a simple analytic expression (Equation~\ref{eq:remnant_mass_L}).

Partial disruptions also play a significant, but previously neglected, role in the approach of stars to disruption when stellar encounters perturb their angular momentum only weakly (the ``empty loss-cone" or ``diffusive" regime).  They can deflect stars into different parts of phase space, they can whittle down formerly massive stars before they are completely disrupted, and they can transform a main sequence star into a distended remnant that may be totally disrupted in a highly unconventional way upon its next return to pericenter.

\section*{Acknowledgements}
We are grateful to the anonymous referee for comments and suggestions that helped us to improve the paper.
We thank Nicholas Stone and Re'em Sari for helpful discussions. This work was partially supported by NSF grant AST-1715032, Simons Foundation grant 559794 and an advanced ERC grant TReX. S.~C.~N. was supported by the grants NSF AST 1515982, NSF OAC 1515969, and NASA 17-TCAN17-0018, and an appointment to the NASA Postdoctoral Program at the Goddard Space Flight Center administrated by USRA through a contract with NASA.  The authors acknowledge the analysis toolkit matplotlib \citep{Hunter:2007} for making the plots in the paper. This research project (or part of this research project) was conducted using computational resources (and/or scientific computing services) at the Maryland Advanced Research Computing Center (MARCC). The authors would like to thank Stony Brook Research Computing and Cyberinfrastructure, and the Institute for Advanced Computational
Science at Stony Brook University for access to the high-performance
SeaWulf computing system, which was made possible by a $\$1.4$M National Science Foundation grant (\#1531492).

\software{
matplotlib \citep{Hunter:2007}; \mesa \citep{Paxton+2011}; 
\harm  \citep{Noble+2009}.
}

\end{document}